\documentclass[aps,pra,superscriptaddress,twocolumn]{revtex4-2}

\usepackage{amsmath,amssymb,mathtools}
\usepackage{color}
\usepackage{graphicx}
\usepackage{subfigure}
\usepackage[colorlinks=true,linkcolor=red,citecolor=blue,urlcolor=magenta,bookmarks]{hyperref}
\usepackage{multirow,makecell}
\usepackage{textcomp}
\usepackage{wasysym}
\usepackage{ulem}

\usepackage{verbatim}

\usepackage[utf8]{inputenc}
\usepackage[T1]{fontenc}

\def \be {\begin{equation}}
\def \ee {\end{equation}}
\def \ba {\begin{array}}
\def \ea {\end{array}}
\def \bea{\begin{eqnarray}}
\def \eea{\end{eqnarray}}
\def \nn {\nonumber}

\def \a {\alpha}
\def \b {\beta}
\def \g {\gamma}

\def \G {\Gamma}
\def \d {\delta}

\def \ve {\varepsilon}
\def \m {\mu}
\def \n {\nu}

\def \l {\lambda}
\def \lam {\lambda}

\def \s {\sigma}

\def \r {\rho}

\def \th {\theta}

\def \f {\frac}

\def \lt {\left}
\def \rt {\right}

\def \sr {\sqrt}
\def \td {\tilde}

\def \pp {\propto}
\def \inf {\infty}

\def \lag {\langle}
\def \rag {\rangle}

\def \ep {\mathrm{e}}
\def \ii {\mathrm{i}}

\def \arctanh {\mathop{\rm arctanh}}

\def \tr {\textrm{tr}}

\def \diag {\mathop{\textrm{diag}}}

\def \and {{~\textrm{and}~}}

\begin{document}

\title{Trace distance between fermionic Gaussian states from a truncation method}

\author{Jiaju Zhang}
\email{jiajuzhang@tju.edu.cn}
\affiliation{Center for Joint Quantum Studies and Department of Physics, School of Science, Tianjin University,\\135 Yaguan Road, Tianjin 300350, China}

\author{M. A. Rajabpour}
\email{mohammadali.rajabpour@gmail.com}
\affiliation{Instituto de Fisica, Universidade Federal Fluminense,\\
      Av.\ Gal.\ Milton Tavares de Souza s/n, Gragoat\'a, 24210-346, Niter\'oi, RJ, Brazil}

\begin{abstract}

  In this paper, we propose a novel truncation method for determining the trace distance between two Gaussian states in fermionic systems. For two fermionic Gaussian states, characterized by their correlation matrices, we consider the von Neumann entropies and dissimilarities between their correlation matrices and truncate the correlation matrices to facilitate trace distance calculations. Our method exhibits notable efficacy in two distinct scenarios. In the first scenario, the states have small von Neumann entropies, indicating finite or logarithmic-law entropy, while their correlation matrices display near-commuting behavior, characterized by a finite or gradual nonlinear increase in the trace norm of the correlation matrix commutator relative to the system size. The second scenario encompasses situations where the two states are nearly orthogonal, with a maximal canonical value difference approaching 2. To evaluate the performance of our method, we apply it to various compelling examples. Notably, we successfully compute the subsystem trace distances between low-lying eigenstates of Ising and XX spin chains, even for significantly large subsystem sizes. This is in stark contrast to existing literature, where subsystem trace distances are limited to subsystems of approximately ten sites. With our truncation method, we extend the analysis to subsystems comprising several hundred sites, thus expanding the scope of research in this field.

\end{abstract}

\maketitle

\section{Introduction}

Quantitative differentiation of quantum states is essential in quantum information theory \cite{Nielsen:2010oan,Watrous:2018rgz}, and also plays important roles in quantum many-body systems, quantum field theories and gravity \cite{Fagotti:2013jzu,Cardy:2014rqa,Lashkari:2014yva,Lashkari:2015dia,Jafferis:2015del,Dong:2016eik,%
Sarosi:2016oks,Lashkari:2016vgj,Sarosi:2016atx,Arias:2016nip,Dymarsky:2016ntg,He:2017vyf,Basu:2017kzo,%
Arias:2017dda,He:2017txy,Suzuki:2019xdq,Zhang:2019kwu,Kusuki:2019hcg,Zhang:2019wqo,Mendes-Santos:2019tmf,Zhang:2019itb,Zhang:2020mjv,%
Arias:2020sgz,deBoer:2020snb,Zhang:2020ouz,Zhang:2020txb,Yang:2021enf,Kudler-Flam:2021rpr,Chen:2021pls,Capizzi:2021zga,Kudler-Flam:2021alo,%
Zhang:2022tgu}.
Various quantities, such as relative entropy, Schatten distances, trace distance, and fidelity, can be employed for this purpose, and in this study, our focus is specifically on the trace distance.
The trace distance, denoted by $D(\rho, \sigma)$, quantifies the distinguishability between two density matrices $\rho$ and $\sigma$ and is defined as half of the trace norm of their difference \cite{Nielsen:2010oan,Watrous:2018rgz}, i.e.,
\be
D(\rho, \sigma) = \frac{1}{2}\mathrm{tr}|\rho - \sigma|.
\ee
In the context of extended quantum systems, the trace distance offers distinct advantages compared to other measures of distinguishability. It is not only a well-defined mathematical distance, but in the scaling limit, it can discern states that remain indistinguishable using other measures \cite{Zhang:2020mjv}.
Furthermore, the trace distance $D$, defined as $D = D(\rho, \sigma)$ between two states $\rho$ and $\sigma$, serves as an upper bound for the difference between their von Neumann entropies $S(\rho) = -\mathrm{tr}(\rho\log\rho)$ and $S(\sigma) = -\mathrm{tr}(\sigma\log\sigma)$, as established by the Fannes-Audenaert inequality \cite{Fannes1973,Audenaert:2006}
\be
|S(\rho) - S(\sigma)| \leq D \log (d-1) - D\log D - (1-D)\log(1-D),
\ee
where $d$ represents the dimension of the Hilbert space.
By employing a specific case of H\"older's inequality, the trace distance also provides an upper bound on the difference in expectation values of an operator
\be
| \mathrm{tr} [ (\rho-\sigma)\mathcal{O} ] | \leq 2 s_{\mathrm{max}}(\mathcal{O}) D(\rho,\sigma),
\ee
where $s_{\mathrm{max}}$ denotes the maximal singular value of the operator $\mathcal{O}$. This property of the trace distance proves highly valuable for defining the eigenstate thermalization hypothesis (ETH), initially proposed in terms of local operator expectation values \cite{Deutsch:1991,Srednicki:1994,Deutsch:2018thx}, and subsequently generalized to the subsystem ETH in relation to the subsystem trace distance \cite{Lashkari:2016vgj,Dymarsky:2016ntg}.
Moreover, the average subsystem trace distance of neighboring states in the spectrum can also be utilized as a novel signature to differentiate between chaotic and integrable many-body quantum systems \cite{Khasseh:2023kxw}.
However, calculating the trace distance for large systems poses a notorious challenge due to the exponential growth of the Hilbert space dimension with the number of qubits.

In this study, we also examine the fidelity, denoted by $F(\rho,\sigma)$, which serves as a benchmark for comparison with the trace distance. The fidelity provides both an upper and a lower bound for the trace distance \cite{Nielsen:2010oan,Watrous:2018rgz}, given by
\be \label{uandlbounds}
1 - F(\rho,\sigma) \leq D(\rho,\sigma) \leq \sqrt{1 - F(\rho,\sigma)^2}.
\ee
For fermionic Gaussian states $\rho_{\Gamma_1}$ and $\rho_{\Gamma_2}$, determined by their respective correlation matrices $\Gamma_1$ and $\Gamma_2$, the exact fidelity can be calculated as \cite{Banchi:2013uht}
\bea \label{FrG1rG2}
&& F(\r_{\G_1},\r_{\G_2}) =
\Big(\det\f{1-\G_1}{2}\Big)^{1/4}
\Big(\det\f{1-\G_2}{2}\Big)^{1/4} \nn\\
&& \hspace{4mm} \times \Big[\det\Big(1+\sr{\sr{\f{1+\G_1}{1-\G_1}}\f{1+\G_2}{1-\G_2}\sr{\f{1+\G_1}{1-\G_1}}}\Big)\Big]^{1/2}.
\eea
It is important to note that a fermionic Gaussian state $\rho$ and its corresponding correlation matrix $\Gamma$ are related as described in equation (\ref{rhoandGamma}).
For recent advancements in calculating the trace distance and fidelity using variational techniques and quantum computers, refer to \cite{Coles:2019kdj,Cerezo:2019tuq,Chen:2020zpo,Li:2021jiv,Agarwal:2021yol,Wang:2023atq}.

In scenarios where the dimension of the Hilbert space is excessively large, exact evaluation of the trace distance becomes impractical, despite the fact that the dominant contributions to the trace distance often arise from a significantly smaller subspace within the Hilbert space. This serves as motivation for our development of a truncation method designed to calculate the trace distance between two Gaussian states in fermionic systems. While straightforward for two pure states, the task becomes nontrivial when dealing with two mixed states, which can manifest as either density matrices of the entire system or reduced density matrices (RDMs) of a subsystem. An ideal situation for the effectiveness of the truncation method can be characterized as follows: Consider two states represented by density matrices expressed as
\be
\rho_1 = \widetilde{\rho} \otimes \widetilde{\rho}_1, ~ \rho_2 = \widetilde{\rho} \otimes \widetilde{\rho}_2,
\ee
where the dimension of $\widetilde{\rho}$ is significantly larger than that of $\widetilde{\rho}_1$ and $\widetilde{\rho}_2$. In this ideal scenario, the specific form of $\widetilde{\rho}$ does not impact the trace distance calculation, leading to a simplified expression:
\be
D(\rho_1, \rho_2) = D(\widetilde{\rho}_1, \widetilde{\rho}_2).
\ee
For more general cases involving two distinct states, $\rho_1$ and $\rho_2$, the objective of the truncation method is to identify $\widetilde{\rho}_1$ and $\widetilde{\rho}_2$ with significantly smaller dimensions compared to $\rho_1$ and $\rho_2$, enabling an approximation
\be
D(\rho_1, \rho_2) \approx D(\widetilde{\rho}_1, \widetilde{\rho}_2).
\ee

A fermionic Gaussian state can be uniquely determined by the two-point correlation functions of the Majorana modes in the system or subsystem, which can be organized to construct a purely imaginary anti-symmetric correlation matrix \cite{Vidal:2002rm,Latorre:2003kg}.
By employing an orthogonal transformation, the correlation matrix can be brought into canonical form. The canonical values of the matrix govern the von Neumann entropy of the state, while the corresponding canonical vectors determine the effective modes. To truncate the Hilbert space dimension, we employ different strategies to select a limited number of effective modes.
The first strategy, which we call the maximal entropy strategy, involves choosing the effective modes with the smallest canonical values for the two relevant states. These modes make the largest contributions to the sum of the von Neumann entropies.
In the second strategy, we compute the canonical values and canonical vectors of the difference between the two correlation matrices and select the effective modes with the largest canonical values. These chosen modes contribute the most to the differences in the two-point correlation functions of the two states, leading us to name this method the maximal difference strategy.
Additionally, we adopt a mixed strategy that combines elements of both the maximal entropy strategy and the maximal difference strategy. In this approach, some modes are chosen according to one strategy, while the remaining modes are selected using the other strategy. We refer to this combined method as the mixed strategy.
It is worth noting that the mixed strategy of the truncation method consistently provides the most accurate estimation of the trace distance due to the contractive nature of the trace distance under partial trace.

The truncation method demonstrates its effectiveness in two intriguing scenarios.
The first scenario encompasses the cases where the states have a low von Neumann entropy, indicating either finite or logarithmic-law entropy, and their correlation matrices exhibit near-commuting behavior, which is characterized by a finite or slow nonlinear increase of the trace norm of the correlation matrix commutator with respect to the system size.
The second scenario includes the situations where the states are nearly orthogonal, with a correlation matrix difference featuring a maximal canonical value approaching 2.
We validate the efficacy of the truncation method through multiple examples, including the calculation of eigenstate RDMs in Ising chains, XX chains, and ground state RDMs in Ising chains with different transverse fields. Notably, we apply the method to compute subsystem trace distances between low-lying eigenstates in critical Ising and XX spin chains, with significantly larger subsystem sizes than those considered in previous works \cite{Zhang:2019wqo, Zhang:2019itb, Zhang:2022tgu}. The previous studies only obtained trace distances for relatively small subsystem sizes, with a maximum size of 7 in \cite{Zhang:2019wqo, Zhang:2019itb} and 12 in \cite{Zhang:2022tgu}. In contrast, this paper presents results with a maximal subsystem size of 359, surpassing the limitations of previous studies and enabling trace distance calculations for much larger subsystem sizes.

The paper is structured as follows:
Section~\ref{sectionTCCMM} presents a detailed explanation of the truncated canonicalized correlation matrix method for fermionic Gaussian states.
The conditions under which the truncation method is effective are discussed in Section~\ref{sectionConditions}.
Examples of its application to eigenstate RDMs in the Ising chain and ground state RDMs in Ising chains with different transverse fields are examined in Sections~\ref{sectionIsing} and \ref{sectionIsingGS}, respectively.
The truncation method is further validated through examples of low-lying eigenstate RDMs in the critical Ising chain (Section~\ref{sectioncriticalIsing}) and half-filled XX chain (Section~\ref{sectionlowlyingXX}).
Concluding discussions are provided in Section~\ref{sectionConclusion}.
Additionally, Appendix~\ref{appendixToy} presents a simple example illustrating the impossibility of a finite truncation method for the trace distance between two orthogonal Gaussian states in the scaling limit.
Appendix~\ref{sectionTDCMM} introduces the truncated diagonalized correlation matrix method, a specialized version of the truncation method, applicable to Gaussian states in the free fermionic theory where the number of excited Dirac modes is conserved.

\section{Truncated canonicalized correlation matrix method}\label{sectionTCCMM}

In this section, we begin by providing a brief overview of the canonicalized correlation matrix method presented in \cite{Zhang:2022tgu}, which is an exact approach requiring the explicit construction of density matrices. However, its efficiency is limited to cases involving very small system sizes. To address this limitation, we introduce the truncated canonicalized correlation matrix method, which allows for proper truncation under specific conditions. By constructing effective density matrices within a significantly smaller subspace of the full Hilbert space, we can approximate the trace distance for cases involving much larger system sizes.

\subsection{Canonicalized correlation matrix method}

We consider a system consisting of $\ell$ spinless fermions, denoted by $a_j$ and $a_j^\dag$ with $j=1,2,\cdots,\ell$. This system can represent either the entire system or a subsystem. To facilitate our analysis, we introduce the Majorana modes defined as
\be
d_{2j-1} = a_j+a_j^\dag, ~
d_{2j} = \ii(a_j-a_j^\dag), ~
j=1,2,\cdots,\ell.
\ee
These Majorana modes satisfy the anticommutation relations:
\be
\{ d_{m_1}, d_{m_2} \} = 2 \d_{m_1m_2}, ~ m_1,m_2=1,2,\cdots,2\ell.
\ee
The system is described by a general Gaussian state with a density matrix $\r$. This state is characterized by a two-point correlation function matrix $\G$ with elements given by
\bea \label{Gammadef}
&& \G_{m_1m_2} = \lag d_{m_1} d_{m_2} \rag_\r - \d_{m_1m_2}, \nn \\
&& m_1,m_2=1,2,\cdots,2\ell.
\eea
All other correlation functions in the Gaussian state can be derived from the correlation matrix $\G$ using Wick contractions.

The correlation matrix $\G$ possesses the properties of being purely imaginary and anti-symmetric, and it can be converted to the canonical form using the approach described in \cite{Hua:1944cah,Becker:1973qv}. In this paper, we adopt the procedure outlined in \cite{Becker:1973qv}. Specifically, we have the equations
\be
\G  u_j = - \ii \g_j v_j, ~
\G  v_j = \ii \g_j u_j, ~ j=1,2,\cdots,\ell,
\ee
where $\g_j$ represents real numbers within the range $[0,1]$ and $u_j$ and $v_j$ are real $2\ell$-component vectors. Each index $j=1,2,\cdots,\ell$ corresponds to a canonical value of $\G$, and the associated vectors $u_j$ and $v_j$ are referred to as the canonical vectors. It is important to note that a $2\ell\times2\ell$ correlation matrix has $\ell$ canonical values and $2\ell$ canonical vectors. The canonical vectors satisfy orthonormality conditions given by
\bea
&& u_{j_1}^T u_{j_2} = v_{j_1}^T v_{j_2} = \d_{j_1j_2}, ~
   u_{j_1}^T v_{j_2} =0, \nn\\
&& j_1,j_2 = 1,2,\cdots,\ell.
\eea
To transform $\G$ into the canonical form, we define the $2\ell\times2\ell$ orthogonal matrix as
\be
Q = (u_1,v_1,u_2,v_2,\cdots,u_\ell,v_\ell),
\ee
where each vector is represented as a column vector. The matrix $Q$ facilitates the transformation of $\G$ according to the equation
\be \label{QTGAKQ}
Q^T \G  Q= \bigoplus_{j=1}^\ell \lt( \ba{cc} 0 & \ii \g_j \\ - \ii \g_j & 0 \ea \rt).
\ee

The density matrix of the Gaussian state can be expressed in terms of the modular Hamiltonian as \cite{Peschel:2002jhw,Barthel:2006wht,Fagotti:2010yr}
\be \label{rhoandGamma}
\r  = \sqrt{\det\f{1-\G }{2}} \exp \Big( -\f12 \sum_{m_1,m_2=1}^{2\ell} W_{m_1m_2} d_{m_1} d_{m_2} \Big),
\ee
where the matrix $W$ is defined as
\be
W  = \arctanh \G .
\ee
Similar to $\G$, the matrix $W$ is also purely imaginary and anti-symmetric and can be transformed as
\bea
&& Q^T W  Q= \bigoplus_{j=1}^\ell \lt( \ba{cc} 0 & \ii \d_j \\ - \ii \d_j & 0 \ea \rt), \nn\\
&& \d_j = \arctanh \g_j, ~
j=1,2,\cdots,\ell.
\eea
We introduce the effective Majorana modes $\td d_{m_1}$, defined as
\be
\td d_{m_1} \equiv \sum_{m_2=1}^{2\ell} Q_{m_2m_1} d_{m_2} , ~
m_1=1,2,\cdots,2\ell,
\ee
which also satisfy the anticommutation relations
\be
\{ \td d_{m_1}, \td d_{m_2} \} = 2 \d_{m_1m_2}, ~ m_1,m_2=1,2,\cdots,2\ell.
\ee
The $2\ell$ effective Majorana modes $\td d_m$ with $m=1,2,\cdots,2\ell$ can be organized into $\ell$ pairs $\{\td d_{2j-1},\td d_{2j}\}$ with $j=1,2,\cdots,\ell$. In the Gaussian state $\r$ characterized by the correlation matrix $\G$, each pair of effective Majorana modes $\{\td d_{2j-1},\td d_{2j}\}$ with $j=1,2,\cdots,\ell$ decouple from the other effective Majorana modes.

The density matrix can be expressed as
\bea \label{rhoofccmm}
&& \r = \Bigg[ \prod_{j=1}^\ell \f{\sqrt{(1+\g_j)(1-\g_j)}}{2} \Bigg]
      \exp \Big( - \ii \sum_{j=1}^\ell \d_j \td d_{2j-1} \td d_{2j} \Big)\nn\\
&& \phantom{\r}  = \prod_{j=1}^\ell \f{1-\ii\g_j\td d_{2j-1}\td d_{2j}}{2}.
\eea
To verify the properties of the density matrix, we check that $\tr\r=1$ and $\tr(\r \td d_{2j-1} \td d_{2j})=\ii \g_j$ for $j=1,2,\cdots,\ell$.
For calculating the trace distance between two Gaussian states, we use their explicit density matrices.
To calculate the fidelity, we can utilize the formula
\bea
&& \sqrt{\r } = \prod_{j=1}^\ell \big\{ [(\sqrt{1+\g_j}+\sqrt{1-\g_j}) \\
&& \phantom{\sqrt{\r} =}-\ii(\sqrt{1+\g_j}-\sqrt{1-\g_j})\td d_{2j-1}\td d_{2j}]/(2\sqrt{2})\big\}, \nn
\eea
which is derived by considering
\be
\sqrt{\r } = \prod_{j=1}^\ell ( \td \a_j + \td \b_j \td d_{2j-1} + \td \g_j \td d_{2j} + \td \d_j \td d_{2j-1} \td d_{2j} ),
\ee
and determining the constants $\td\a_j$, $\td\b_j$, $\td\g_j$, and $\td\d_j$ by comparing $\sqrt{\r}^2$ with $\r$ in (\ref{rhoofccmm}).

\subsection{Truncated canonicalized correlation matrix method}

To calculate the approximate trace distance and fidelity, we employ three strategies for implementing the truncated canonicalized correlation matrix method. The approach involves dimension truncation of the Hilbert space to a subspace that is spanned by a properly selected, limited number of effective Majorana modes.

\subsubsection{Maximal entropy strategy} \label{MESofTCCMM}

\begin{figure}[ht]
  \centering
  \includegraphics[width=0.37\textwidth]{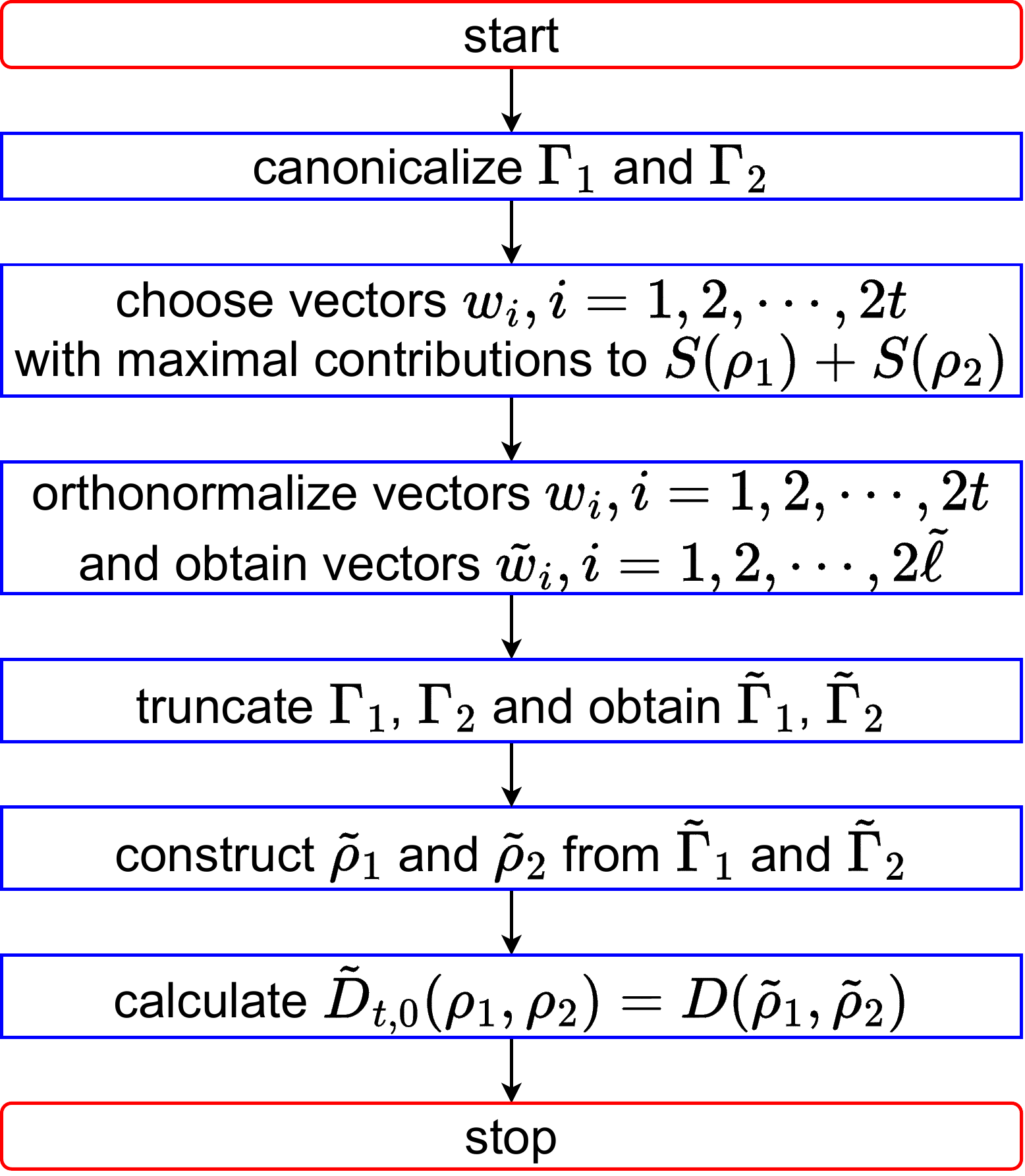}\\
  \caption{The flowchart illustrates the evaluation of the trace distance using the maximal entropy strategy within the truncated canonicalized correlation matrix method.}
  \label{TCCMM-MES}
\end{figure}

Using a similarity transformation, we can change the corresponding density matrix $\r$ into a specific form for the correlation matrix $\G$ (as given by Equation (\ref{QTGAKQ})). This transformation is described in references \cite{Vidal:2002rm,Latorre:2003kg} and yields
\be
\r  \cong \bigotimes_{j=1}^\ell \lt( \ba{cc} \f{1-\g_j}{2} & \\ & \f{1+\g_j}{2} \ea \rt).
\ee
The von Neumann entropy of $\r$ is simply the sum of the Shannon entropies of each effective probability distribution $\{\f{1-\g_j}{2}, \f{1+\g_j}{2}\}$ with $j=1,2,\cdots,\ell$
\be
S(\r)  = \sum_{j=1}^\ell \Big( -\f{1-\g_j}{2}\log\f{1-\g_j}{2} -\f{1+\g_j}{2}\log\f{1+\g_j}{2} \Big).
\ee
For low-rank states $\r$, the entropy $S(\r)$ is not large, allowing us to truncate the effective probability distribution $\{\f{1-\g_j}{2}, \f{1+\g_j}{2}\}$ with $j=1,2,\cdots,\ell$ to only include a few values with the smallest $\g_j$.
Applying the same idea to two Gaussian states $\r_1$ and $\r_2$, we truncate the density matrices and calculate their trace distance. It's important to note that for high-rank states $\r$, where the entropy $S(\r)$ is large, imposing truncation is generally inefficient and the maximal entropy strategy introduced in this subsection is expected to fail.
The flowchart in Figure \ref{TCCMM-MES} illustrates the evaluation of the trace distance using the maximal entropy strategy of the truncated canonicalized correlation matrix method.

We consider two general Gaussian states $\r_1$ and $\r_2$ with correlation matrices $\G_1$ and $\G_2$. The canonical values and vectors $(\g_j,u_j,v_j)$ with $j=1,2,\cdots,\ell$ correspond to the matrix $\G_1$, while the canonical values and vectors $(\g_j,u_j,v_j)$ with $j=\ell+1,\ell+2,\cdots,2\ell$ correspond to the matrix $\G_2$.
To choose $2t$ effective Majorana modes whose corresponding canonical values contribute the most to the entropy sum of the two states, given by
\bea \label{entropysum}
&& S(\r_1) + S(\r_2)  = \sum_{j=1}^{2\ell} \Big( -\f{1-\g_j}{2}\log\f{1-\g_j}{2} \nn\\ && \phantom{S(\r_1) + S(\r_2)  =} -\f{1+\g_j}{2}\log\f{1+\g_j}{2} \Big),
\eea
we sort the $2\ell$ sets of canonical values and vectors $(\g_j,u_j,v_j)$ with $j=1,2,\cdots,2\ell$ in ascending order based on the values of $\g_j\in[0,1]$. Then, we select the first $2t$ canonical vectors $(u_j,v_j)$ corresponding to the first $t$ canonical values $\g_j$, and denote these selected vectors as $w_i$, where $i=1,2,\cdots,2t$.

The $2t$ real vectors $w_i$ with $i=1,2,\cdots,2t$ are not generally orthogonal and may not be independent. To orthogonalize and normalize, a.k.a.\ orthonormalize, these vectors, we diagonalize the $2t \times 2t$ matrix $V$, whose entries are given by $V_{i_1i_2} = w_{i_1}^T w_{i_2}$ for $i_1,i_2=1,2,\cdots,2t$. If the $2t$ vectors $w_i$ are not independent, the $2t \times 2t$ matrix $V$ will have eigenvalues close to zero. It's important to consider numerical errors that can lead to very small eigenvalues. We sort the eigenvalues and orthonormal eigenvectors $(\a_i,a_i)$ with $i=1,2,\cdots,2t$ of $V$ based on the values of $\a_i$ in descending order, discarding those smaller than a certain cutoff (e.g., $10^{-9}$). In cases where the number of remaining vectors is odd, we add an extra vector with the largest eigenvalue smaller than the cutoff. After discarding, we obtain $2\td \ell$ sets of eigenvalues and eigenvectors $(\a_i,a_i)$ with $i=1,2,\cdots,2\td\ell$, from which we define the $2\ell$-component orthonormal vectors as
\be
\td w_i \equiv \f{1}{\sqrt{\a_i}} \sum_{i'=1}^{2t} [a_i]_{i'} w_{i'} , ~ i=1,2,\cdots,2\td\ell.
\ee
It can be easily verified that the orthonormality condition $\td w_{i_1}^T \td w_{i_2} = \d_{i_1i_2}$ holds for $i_1,i_2=1,2,\cdots,2\td\ell$. We then define the $2\ell\times2\td\ell$ matrix $\td Q = ( \td w_1, \td w_2,\cdots, \td w_{2\td\ell} )$, and the $2\td\ell \times 2\td\ell$ truncated correlation matrices $\td \G_1 = \td Q^T \G_1 \td Q$ and $\td \G_2 = \td Q^T \G_2 \td Q$.

Using the truncated correlation matrices $\td \G_1$ and $\td \G_2$, we construct the $2^{\td\ell}\times 2^{\td\ell}$ truncated density matrices $\td \r_1$ and $\td \r_2$ using the canonicalized correlation matrix method explained in the previous subsection. Subsequently, we calculate the approximate trace distance and fidelity as
\bea \label{tdDt0tdFt0}
&& \td D_{t,0}(\r_1,\r_2) \equiv D(\td \r_1,\td \r_2), \\
&& \td F_{t,0}(\r_1,\r_2) \equiv F(\td \r_1,\td \r_2).
\eea
To facilitate later discussions, we introduced the subscript $t,0$ in these equations for convenience.

\subsubsection{Maximal difference strategy} \label{MDSofTCCMM}

\begin{figure}[ht]
  \centering
  \includegraphics[width=0.3\textwidth]{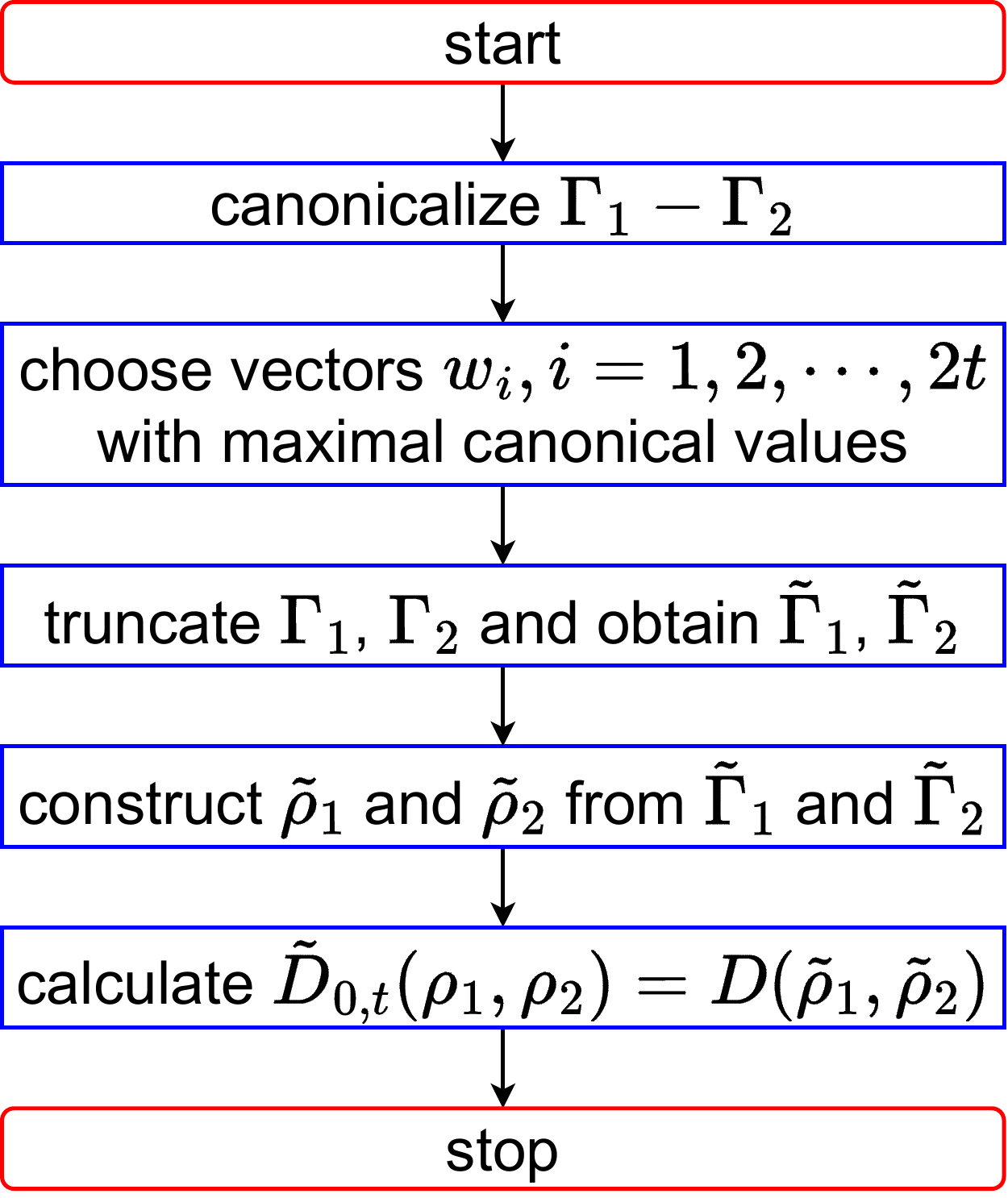}\\
  \caption{The flowchart depicts the calculation of the trace distance using the maximal difference strategy within the truncated canonicalized correlation matrix method.}
  \label{TCCMM-MDS}
\end{figure}

The entries of the correlation matrix difference between two states are simply the differences of their two-point functions. To truncate the Hilbert space, we can select the effective Majorana modes that maximize the differences in two-point functions among them. Both the differences in two-point functions and the trace distance provide measures of the difference between the two states. In certain cases, we expect that the modes with the maximum differences in two-point functions contribute the most to the trace distance. In this subsection, we introduce the maximal difference strategy for the truncation method based on the canonicalization of the correlation matrix difference. The flowchart in Figure \ref{TCCMM-MDS} illustrates the calculation of the trace distance using the maximal difference strategy within the truncated canonicalized correlation matrix method.

For two fermionic Gaussian states, denoted by $\r_1$ and $\r_2$, we canonicalize the correlation matrix difference $\G_1-\G_2$ and sort the canonical values and canonical vectors $(\g_j,u_j,v_j)$ with $j=1,2,\cdots,\ell$ in descending order based on the canonical values. We rename the $2t$ canonical vectors $(u_j,v_j)$ with $j=1,2,\cdots,t$ corresponding to the first $t$ canonical values $\g_j$ as $w_i$ with $i=1,2,\cdots,2t$. It is important to note that for each canonical value, there are two corresponding canonical vectors. These $2t$ vectors $w_i$ with $i=1,2,\cdots,2t$ are already orthonormal. We construct the $2\ell \times 2t$ matrix
\be
\td Q = ( w_1, w_2, \cdots, w_{2t} ),
\ee
and obtain the $2t \times 2t$ truncated correlation matrices
\be
\td \G_1 = \td Q^T \G_1 \td Q, ~ \td \G_2 = \td Q^T \G_2 \td Q.
\ee
Using the canonicalized correlation matrix method, we construct the $2^{t} \times 2^{t}$ truncated density matrices $\td \r_1$ and $\td \r_2$ from the $2t \times 2t$ truncated correlation matrices $\td \G_1$ and $\td \G_2$. Finally, we calculate the approximate trace distance and fidelity using the truncated density matrices as
\bea
&& \td D_{0,t}(\r_1,\r_2) = D(\td \r_1,\td \r_2), \\
&& \td F_{0,t}(\r_1,\r_2) = F(\td \r_1,\td \r_2).
\eea
The subscript $0,t$ is used for convenience in later discussions.

\subsubsection{Mixed strategy} \label{MSofTCCMM}

\begin{figure}[ht]
  \centering
  \includegraphics[width=0.46\textwidth]{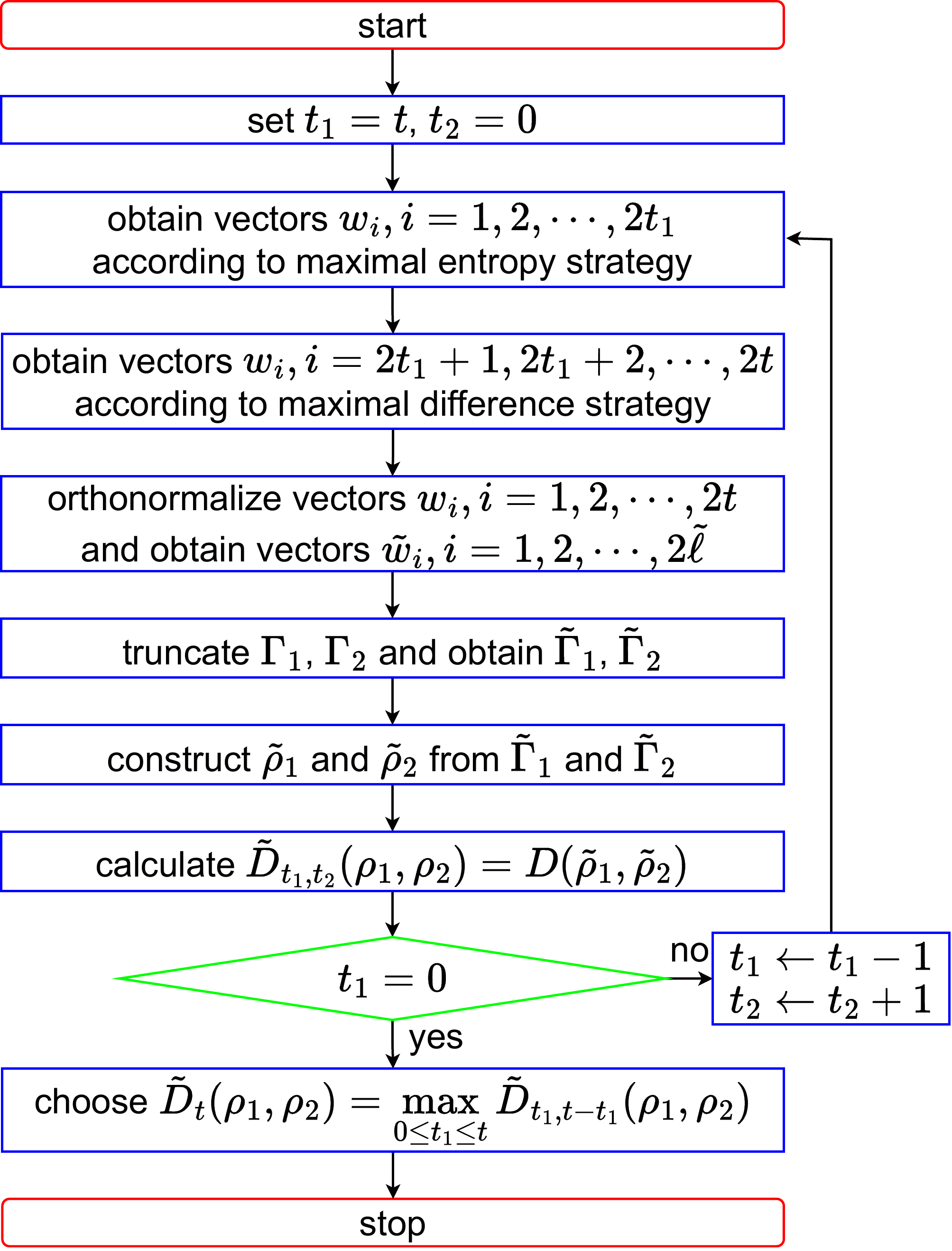}\\
  \caption{The flowchart illustrates the calculation of the trace distance using the mixed strategy within the truncated canonicalized correlation matrix method.}
  \label{TCCMM-MS}
\end{figure}

Given two fermionic Gaussian states, it is not guaranteed that either the maximal entropy strategy or the maximal difference strategy will work in all cases. The essence of the truncation method lies in selecting effective Majorana modes to truncate the Hilbert space. In this subsection, we introduce a mixed strategy for the truncation method, where some effective Majorana modes are chosen based on the maximal entropy strategy and others are chosen based on the maximal difference strategy. The flowchart illustrating this mixed strategy is shown in Figure \ref{TCCMM-MS}.

As before, we consider two fermionic Gaussian states, denoted by $\r_1$ and $\r_2$, with correlation matrices $\G_1$ and $\G_2$. For a fixed truncation number $t$, we split it into $t=t_1+t_2$. Firstly, we obtain $2t_1$ vectors $w_i$ with $i=1,2,\cdots,2t_1$ using the maximal entropy strategy. Next, we obtain $2t_2$ vectors $w_i$ with $i=2t_1+1,2t_1+2,\cdots,2t$ using the maximal difference strategy. Similar to the maximal entropy strategy, the $2t$ vectors $w_i$ with $i=1,2,\cdots,2t$ are generally not orthogonal and may not be independent. The procedure for orthonormalizing these $2t$ vectors $w_i$ with $i=1,2,\cdots,2t$ is the same as that in the maximal entropy strategy, and we will not repeat the details here. In the end, we obtain the $2\ell$-component orthonormal vectors $\td w_i$ with $i=1,2,\cdots,2\td\ell$, where $\td\ell \leq t$.

Then, we define the $2\ell\times2\td\ell$ matrix
\be \td Q = ( \td w_1, \td w_2,\cdots, \td w_{2\td\ell} ), \ee
and the truncated correlation matrices of size $2\td\ell \times 2\td\ell$ as
\be
\td \G_1 = \td Q^T \G_1 \td Q, ~ \td \G_2 = \td Q^T \G_2 \td Q.
\ee
Using the canonicalized correlation matrix method, we construct the $2^{\td\ell}\times 2^{\td\ell}$ truncated density matrices $\td \r_1$ and $\td \r_2$ from the $2\td\ell\times2\td\ell$ truncated correlation matrices $\td \G_1$ and $\td \G_2$. Finally, we calculate the approximate trace distance and fidelity using the truncated density matrices as
\bea
&& \td D_{t_1,t_2}(\r_1,\r_2) = D(\td \r_1,\td \r_2), \\
&& \td F_{t_1,t_2}(\r_1,\r_2) = F(\td \r_1,\td \r_2).
\eea
When $t_2=0$, the mixed strategy reduces to the maximal entropy strategy, and when $t_1=0$, it reduces to the maximal difference strategy.

For a fixed $t$, there are $t+1$ different ways to split it, resulting in $t+1$ approximate values of the trace distance and fidelity. Since the trace distance is contractive under partial trace and the fidelity is expansive under partial trace, the exact trace distance of the two states is always greater than or equal to the approximate trace distance obtained from the truncation method, and the exact fidelity is always less than or equal to the approximate fidelity obtained from the truncation method. For fixed $t<\ell$, we can obtain the best estimation of the trace distance and fidelity using the mixed truncation strategy as
\bea
&& \td D_{t}(\r_1,\r_2) = \max_{0 \leq t_1 \leq t} \td D_{t_1,t-t_1}(\r_1,\r_2), \\
&& \td F_{t}(\r_1,\r_2) = \min_{0 \leq t_1 \leq t} \td F_{t_1,t-t_1}(\r_1,\r_2).
\eea
When $\ell\leq t$, no truncation is necessary, and we can directly calculate the exact trace distance and fidelity using the canonicalized correlation matrix method.

The mixed strategy of the truncation matrix method consistently provides higher precision than the other two strategies. Hence, when we refer to the truncated canonicalized correlation matrix method without specifying the strategy, we imply the mixed strategy of the truncation method.

Estimating the computational expense for each step of the truncation method is also intriguing.
Within the algorithm, the task of canonicalizing either the correlation matrix or the difference between two correlation matrices takes a polynomial amount of time relative to the system size $\ell$.
Constructing the truncated correlation matrices also requires polynomial time relative to the truncation number $t$.
However, constructing the truncated density matrices and calculating the trace distance between them is an exponential-time task in relation to the truncation number $t$.
Therefore, maintaining a reasonably small truncation number $t$ is crucial to ensure the efficiency of the truncation method.

\section{Conditions for the truncation method to work} \label{sectionConditions}

Before applying the three strategies of the truncation method to concrete examples, we will analyze the conditions under which the truncation method works and when it might not work in this section.

Based on the scaling behavior of von Neumann entropies, we can classify states into two categories: small-entropy states and large-entropy states. Small-entropy states have finite or logarithmic von Neumann entropies with respect to system size in the scaling limit. On the other hand, large-entropy states exhibit volume-law von Neumann entropies in the scaling limit.
The trace distance between two fermionic Gaussian states depends not only on the canonical values of the correlation matrices but also on the canonical vectors. The efficiency of the truncation method relies on the commutativity of the density matrices, or equivalently, the commutativity of the correlation matrices.

In the best scenario, the two correlation matrices commute, denoted by $\G_1 \G_2 = \G_2 \G_1$, and the canonical vectors of the correlation matrices, i.e., the effective Majorana modes of the states in canonical form, are parallel. In this case, one can calculate the trace distance for a relatively large system \cite{Zhang:2020mjv}. To quantify the noncommutativity of general correlation matrices $\G_1$ and $\G_2$, we introduce the noncommutativity trace norm given by
\be \label{NCTN}
N(\G_1,\G_2) = \tr | \G_1 \G_2 - \G_2 \G_1 |.
\ee
In the worst scenario, the noncommutativity trace norm grows linearly $N(\G_1,\G_2) \pp \ell$ in the scaling limit $\ell \to \infty$. In this case, the effective Majorana modes of the two states are significantly different, and the truncation method generally fails to provide accurate results.

For small-entropy states, a sufficient condition for the truncation method to work is that the states nearly commute, meaning the noncommutativity trace norm of their correlation matrices is either constant or grows slowly with the system size. In all the cases of small-entropy states considered in section~\ref{sectionIsing}, the noncommutativity trace norms do not grow faster than $\log \log \log \ell$, and the truncation method works well. However, for the cases discussed in section~\ref{sectionIsingGS}, where the noncommutativity trace norms grow linearly, the truncation method generally fails.

Dealing with large-entropy states poses a more complex situation. In some cases, no finite truncation method is possible, as illustrated in appendix~\ref{appendixToy}. We currently lack a general criterion to determine when the truncation method works for more general fermionic Gaussian states, especially when involving small-entropy states that do not nearly commute or large-entropy states.

However, there is a special case where the maximal difference strategy of the truncation method works. If the maximal canonical value of the correlation matrix difference $\G_1-\G_2$ is exactly $2$, the maximal difference strategy can select this canonical value and its corresponding canonical vectors, resulting in a trace distance $\td D(\r_1,\r_2)=1$. Taking into account the contractive property of the trace distance under partial trace, $D(\r_1,\r_2) \geq \td D(\r_1,\r_2)$, and the fact that $D(\r_1,\r_2) \leq 1$, we obtain the exact trace distance $D(\r_1,\r_2) = 1$. When the maximal canonical value of the correlation matrix difference $\G_1-\G_2$ is nearly $2$, we similarly obtain a well-approximated trace distance $D(\r_1,\r_2) \approx 1$.

Based on the above analysis, we propose the following protocol for using the truncation method:
\begin{widetext}
\be \label{TheConditions}
{\rm two~Gaussian~states}
\lt\{\ba{l}
{\rm small~entropy~and~nearly~commuting~conditions} ~~ {\rm it~works} \\
{\rm otherwise}
     \lt\{ \ba{ll}
     {\rm nearly~orthogonal~condition} & {\rm it~works}          \\
     {\rm otherwise}                   & {\rm it~possibly~fails.}
     \ea \rt.
\ea\rt.
\ee
\end{widetext}
Note that the small entropy condition refers to finite or logarithmic-law entropy for both states in the scaling limit. The nearly commuting condition implies that the noncommutativity trace norm of the two correlation matrices approaches a constant or grows very slowly with the system size. The nearly orthogonal condition represents a special case where the maximal canonical value of the correlation matrix difference is nearly 2. It is worth noting that not all nearly orthogonal Gaussian states have the maximal canonical value of the correlation matrix difference nearly equal to 2, as shown in the example in appendix~\ref{appendixToy}.

When the conditions in (\ref{TheConditions}) are satisfied, it is necessary to determine the required truncation number $t$.
For cases satisfying the small entropy and nearly commuting conditions, we need $t\gtrsim\log\ell$, where $\ell$ is the system size, to ensure that the truncated density matrices can approximately reproduce the von Neumann entropies of the two states and then the trace distance between the two states.
For cases satisfying the nearly orthogonal condition, a finite value of the truncation number $t$ is sufficient to achieve an approximate trace distance.

\section{Examples of eigenstate RDMs in Ising chain} \label{sectionIsing}

We utilize the truncation method to calculate the subsystem trace distance between the ground and excited states in the Ising chain with transverse field. The Hamiltonian of the system, with periodic boundary conditions, is given by
\be \label{HIsing}
H = - \f12 \sum_{j=1}^L ( \s_j^x\s_{j+1}^x + \l \s_j^z ),
\ee
where $\sigma_{j+L}^{x,y,z}=\sigma_j^{x,y,z}$ are the Pauli matrices.
Through the Jordan-Wigner transformation, this Hamiltonian can be rewritten as a fermionic chain \cite{Lieb:1961fr,Katsura:1962hqz,Pfeuty:1970ayt}
\bea
&& H = \sum_{j=1}^L \Big[ \lam \Big( a_j^\dag a_j - \f12 \Big)
                     - \f12 ( a_j^\dag a_{j+1} + a_{j+1}^\dag a_j \nn\\
&& \phantom{H=}                          + a_j^\dag a_{j+1}^\dag + a_{j+1} a_j ) \Big].
\eea
After further Fourier transformation and Bogoliubov transformation, it takes the form
\be \label{DiagonalFermionicChain}
H = \sum_k \ve_k \Big( c_k^\dag c_k - \f12 \Big), ~
\ve_k \equiv \sr{ \l^2 - 2\l \cos \f{2\pi k}{L} + 1 } .
\ee
We focus on the case where $L$ is an even integer and consider the Neveu-Schwarz (NS) sector with anti-periodic boundary conditions for the fermions.
In the NS sector, the momenta take on half-integer values
\be
k = -\f{L-1}{2}, \cdots, -\f12, \f12, \cdots,\f{L-1}{2}.
\ee

The energy eigenstates are characterized by the momenta of the excited quasiparticles $K=\{k_1,k_2,\cdots,k_r\}$, and the RDM of a quasiparticle excited state is also a Gaussian state. For the subsystem $A=[1,\ell]$ in the state $|K\rangle$, we define the $2\ell \times 2\ell$ correlation matrix $\Gamma_{A,K}$, which has entries:
\bea
&& [\G_{A,K}]_{2j_1-1,2j_2-1} = [\G_{A,K}]_{2j_1,2j_2}=f_{j_2-j_1}^K, \nn\\
&& [\G_{A,K}]_{2j_1-1,2j_2} = -[\G_{A,K}]_{2j_2,2j_1-1}=g_{j_2-j_1}^K,
\eea
where $j_1,j_2 = 1,2,\cdots,\ell$. The quantities $f_j^K$ and $g_j^K$ are defined as \cite{Vidal:2002rm,Latorre:2003kg,Alba:2009th,Alcaraz:2011tn,Berganza:2011mh}
\bea
&& f^K_j \equiv  \f{2\ii}{L} \sum_{k \in K}\sin\Big(\f{2\pi j k}{L}\Big), \nn\\
&& g^K_j \equiv -\f{\ii}{L} \sum_{k \notin K} \cos\Big( \f{2\pi j k}{L} -\th_k \Big) \nn\\
&& \phantom{g^K_j \equiv}
          + \f{\ii}{L} \sum_{k \in K} \cos\Big( \f{2\pi j k}{L}-\th_k \Big).
\eea
where the angle $\theta_k$ is determined by
\be \label{epiithk}
\ep^{\ii\th_k} = \f{\l-\cos \f{2\pi k}{L}+\ii\sin \f{2\pi k}{L}}{\sr{ \l^2 - 2\l \cos \f{2\pi k}{L} + 1 }}.
\ee

We compute the approximate subsystem trace distance and fidelity for various states. In the case of a gapped Ising chain, the von Neumann entropies of the RDMs, which are also known as entanglement entropies, exhibit finite values in the scaling limit for few-quasiparticle states \cite{Pizorn:2012aut,Berkovits:2013mii,Molter:2014qsb,Castro-Alvaredo:2018dja,Castro-Alvaredo:2018bij,Zhang:2020ouz,Zhang:2020vtc,Zhang:2020dtd,%
Zhang:2020txb,Zhang:2021bmy}. When the Ising chain is critical, the leading entanglement entropies in few-quasiparticle states follow a logarithmic law.
As for many-quasiparticle states, there are two typical scaling laws observed for the entanglement entropy. In cases where nearly all excited modes occur successively, the entropies adhere to the logarithmic law \cite{Holzhey:1994we,Vidal:2002rm,Latorre:2003kg,Calabrese:2004eu}. We refer to these as logarithmic-law many-quasiparticle states or simply logarithmic-law states. On the other hand, when certain fixed patterns are successively excited in an extended region of momentum space, with a mix of excited and non-excited modes within these patterns, the entropies follow a volume law in the regime $1 \ll \ell \ll L$ \cite{Alba:2009th}. These are referred to as volume-law many-quasiparticle states or volume-law states. It is worth noting that the scaling behavior of the entanglement entropy remains unchanged when a finite number of modes are altered in a given state. To summarize, we can classify the typical states as follows:
\be \label{classification}
\Bigg\{
\ba{l}
\textrm{few-quasiparticle states} \\
\textrm{many-quasiparticle states} ~ 
\bigg\{
\ba{l}
\textrm{logarithmic-law states} \\
\textrm{volume-law states}. \\
\ea
\ea
\ee

Note that the few-quasiparticle states and the logarithmic-law states are called small-entropy states and the volume-law states are referred to as large-entropy states.
Here are some examples of few-quasiparticle states
\bea
&& K_1 = \Big\{ \f12, \f32 \Big\}, \nn\\
&& K_2 = \Big\{ \f12, \f32, \f52, \f72 \Big\},
\eea
logarithmic-law states
\bea
&& K_3 = \Big\{ -\f{L-1}{2}, -\f{L-3}{2}, \cdots, -\f12 \Big\}, \nn\\
&& K_4 = \Big\{ -\f{L-1}{2}, -\f{L-3}{2}, \cdots, \f32 \Big\},
\eea
and volume-law states
\bea
&& K_5 = \Big\{ -\f{L-1}{2}, -\f{L-5}{2}, \cdots, -\f32 \Big\}, \nn\\
&& K_6 = \Big\{ \f12, \f52, \cdots, \f{L-3}{2} \Big\}.
\eea

\begin{figure*}[tp]
  \centering
  \includegraphics[width=0.92\textwidth]{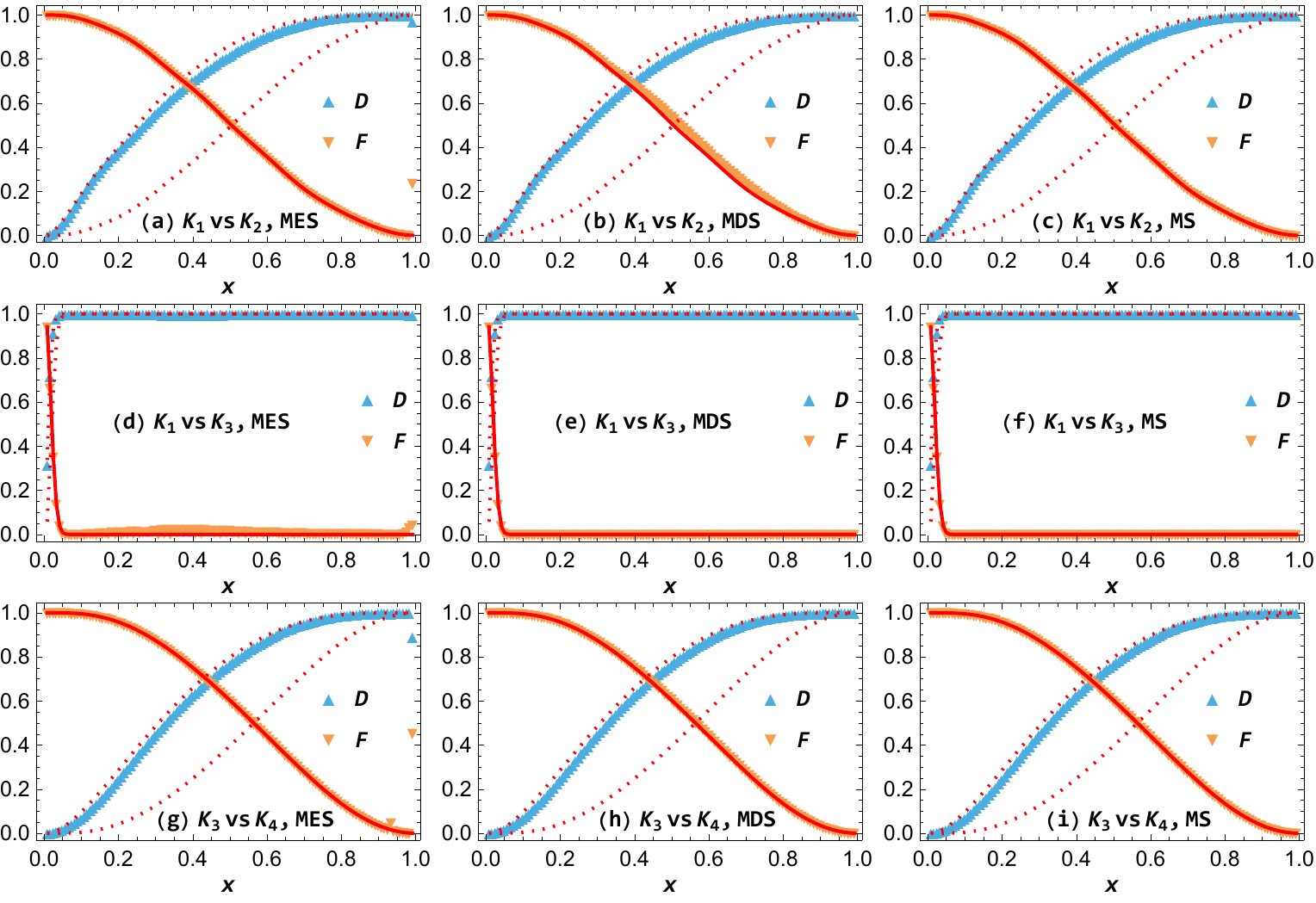}\\
  \caption{The symbols provided exemplify the approximate trace distance ($D$) and fidelity ($F$) between two small-entropy states obtained using the truncated canonicalized correlation matrix method.
  Three strategies are employed: maximal entropy strategy (MES, depicted on the left), maximal difference strategy (MDS, shown in the middle), and the mixed strategy (MS, displayed on the right) in an Ising chain with a transverse field.
  The red solid lines represent exact fidelity results calculated using equation (\ref{FrG1rG2}).
  The red dotted lines represent lower and upper bounds of the trace distance derived from fidelity, as specified by equation (\ref{uandlbounds}).
  In this scenario, we have set $\l=1$, $L=120$, and $t=10$.}
  \label{FigureIsingTDF1}
\end{figure*}

\begin{figure*}[tp]
  \centering
  \includegraphics[width=0.92\textwidth]{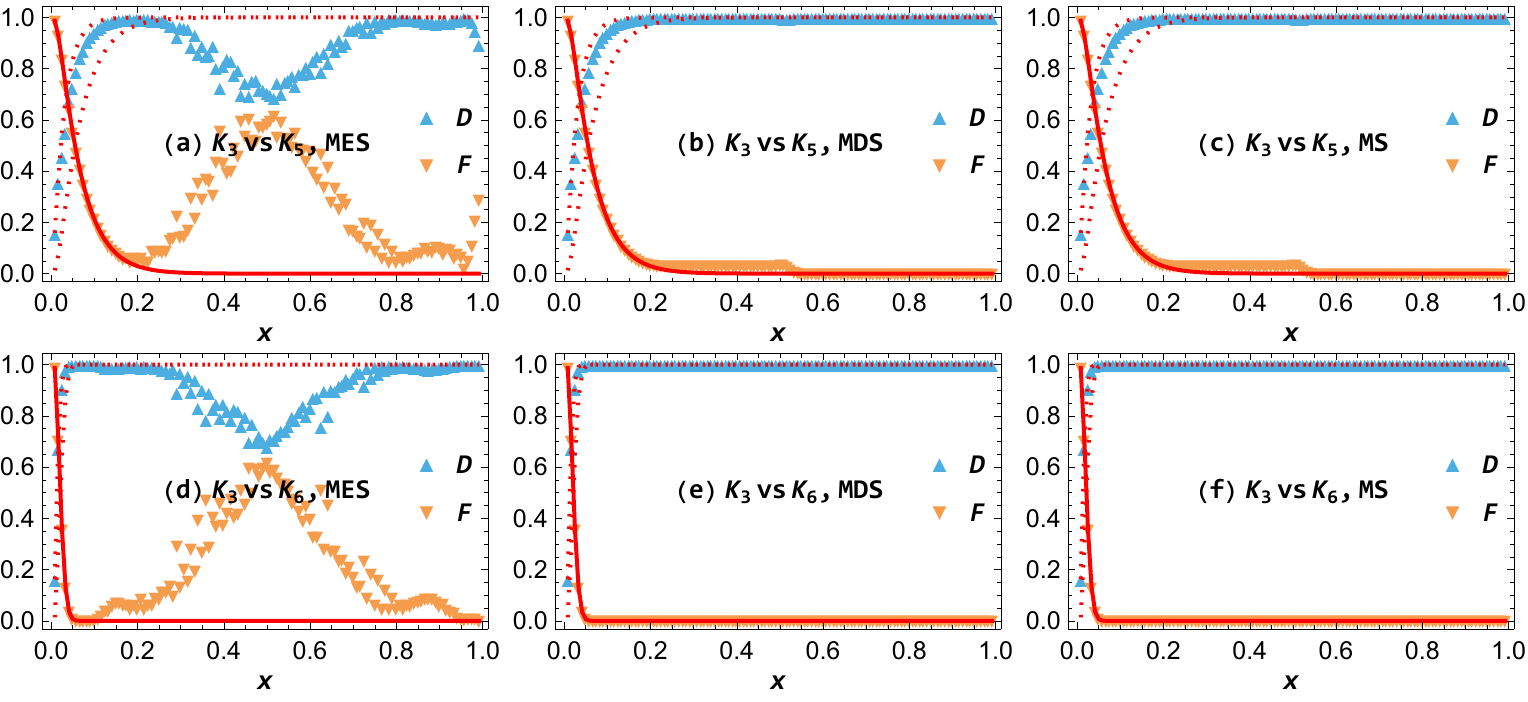}\\
  \caption{The symbols provided demonstrate the approximate trace distance ($D$) and fidelity ($F$) between a small-entropy state and a large-entropy state obtained using the truncation method. Three strategies are employed: maximal entropy strategy (MES, left), maximal difference strategy (MDS, middle), and the mixed strategy (MS, right) in an Ising chain with a transverse field.
  The red solid lines represent exact fidelity results calculated using equation (\ref{FrG1rG2}).
  The red dotted lines represent lower and upper bounds of the trace distance derived from fidelity, as specified by equation (\ref{uandlbounds}).
  We have set $\l=1$, $L=120$, and $t=10$.}
  \label{FigureIsingTDF2}
\end{figure*}

We examine six categories of trace distance and fidelity involving various representative states. These categories include the following: 
(i) two few-quasiparticle states, 
(ii) two logarithmic-law states, 
(iii) two volume-law states, 
(iv) one few-quasiparticle state and one logarithmic-law state, 
(v) one few-quasiparticle state and one volume-law state, and 
(vi) one logarithmic-law state and one volume-law state.
In Figure~\ref{FigureIsingTDF1}, we provide instances of trace distance and fidelity between two small-entropy states. Additionally, Figure~\ref{FigureIsingTDF2} presents examples of trace distance and fidelity between one small-entropy state and one large-entropy state.
Both figures compare the approximate fidelity obtained from the truncated correlation matrix method with the exact results calculated directly from correlation matrix (\ref{FrG1rG2}). It is noteworthy that the exact fidelity establishes lower and upper bounds for the trace distance as per (\ref{uandlbounds}).

Based on the examples presented in Figure~\ref{FigureIsingTDF1} and additional unshown examples, we observe that the mixed strategy of the truncation method performs effectively when dealing with small-entropy states exclusively. We verify that the conditions outlined in (\ref{TheConditions}) are met for these states. Notably, the noncommutativity trace norm approaches a constant or exhibits slow growth with increasing system size in these cases.

The situation becomes more intricate when at least one large-entropy state is involved, as illustrated in Figure~\ref{FigureIsingTDF2}. In all cases where at least one large-entropy state is present and the maximal difference strategy of the truncation method proves successful, there are numerous canonical values of the correlation matrix difference that approach nearly 2. An example of this is seen in the case of $K_3$ versus $K_6$. Similarly, for cases involving only small-entropy states where the maximal difference strategy of the truncation method is effective, there are also considerable numbers of canonical values of the correlation matrix difference that approach nearly 2, as exemplified by the case of $K_1$ versus $K_3$. The maximal difference strategy of the truncation method performs well in these cases, where the trace distance rapidly approaches 1 and the fidelity rapidly decreases to 0 with an increase in subsystem size. This behavior aligns with the conditions specified in (\ref{TheConditions}). In all these instances, there exists a sizable consecutive segment in the momentum space where the modes are either fully excited or unexcited in one state compared to the other.

\begin{figure}[tp]
  \centering
  \includegraphics[width=0.45\textwidth]{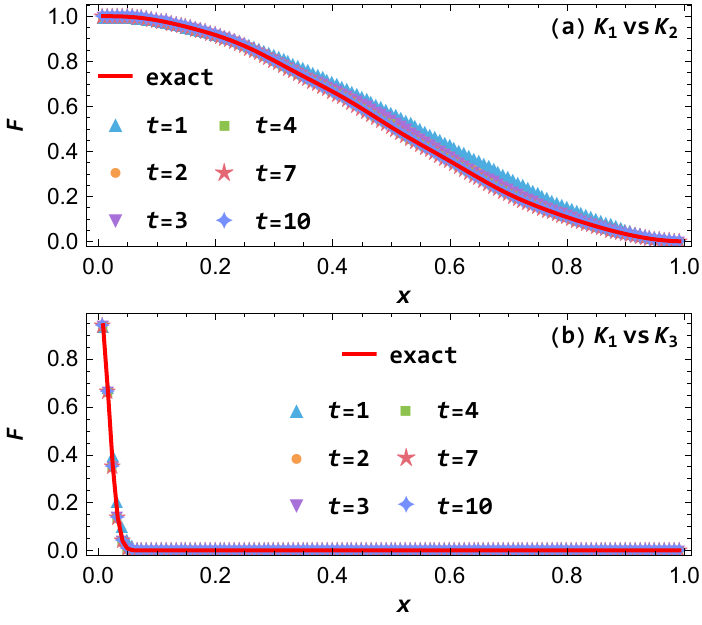}\\
  \caption{The correlation between the truncation number and the approximate fidelity ($F$) is demonstrated for specific examples in the Ising chain, showcasing the effectiveness of the truncation method.
  The solid red lines (exact) represent the exact fidelity calculated using formula (\ref{FrG1rG2}).
  The symbols represent the approximate fidelity obtained using the mixed strategy of the truncation method with different truncation numbers.
  In this analysis, we have set $\l=1$ and $L=120$.}
  \label{FigureIsingFK1vsK2nK1vsK3}
\end{figure}

Furthermore, Figure~\ref{FigureIsingFK1vsK2nK1vsK3} illustrates the relationship between the truncation number and the approximate fidelity specifically for cases where the truncation method proves successful. In the scenario of $K_1$ versus $K_2$, the conditions of small entropy and near-commuting states in (\ref{TheConditions}) are satisfied, requiring the truncation number $t$ to be at least the logarithm of the system size $L$. In Figure~\ref{FigureIsingFK1vsK2nK1vsK3}, where $L=120$, it is necessary to have $t\gtrsim\log L\approx 4.8$. For the case of $K_1$ versus $K_3$, the nearly orthogonal conditions in (\ref{TheConditions}) are satisfied, and even a very small finite value of the truncation number $t$ is sufficient to yield a well-approximated fidelity. These results are consistent with the analysis presented at the end of Section~\ref{sectionConditions}.

\section{Examples of ground state RDMs in different Ising chains}\label{sectionIsingGS}

In this section, we focus on evaluating the trace distance between eigenstate RDMs obtained from different Hamiltonians. Specifically, we calculate the subsystem trace distance among the ground states in the NS sector of Ising chains (\ref{HIsing}) with varying transverse fields $\lambda$. The ground state in the NS sector, denoted by $|G(\lambda)\rangle$, is characterized by the condition
\be
c_k |G(\lambda)\rangle = 0, ~ k=-\frac{L-1}{2}, \cdots, \frac{L-1}{2}.
\ee
The state $|G(\lambda)\rangle$ depends on the transverse field $\lambda$, and for convenience, we will use $|\lambda\rangle \equiv |G(\lambda)\rangle$ to represent the NS sector ground state in the Ising chain with transverse field $\lambda$. Additionally, we introduce the notation $\Gamma_{A,\lambda}\equiv\Gamma_{A,G(\lambda)}$ to denote the subsystem correlation matrix.

\begin{figure*}[tp]
  \centering
  \includegraphics[width=0.92\textwidth]{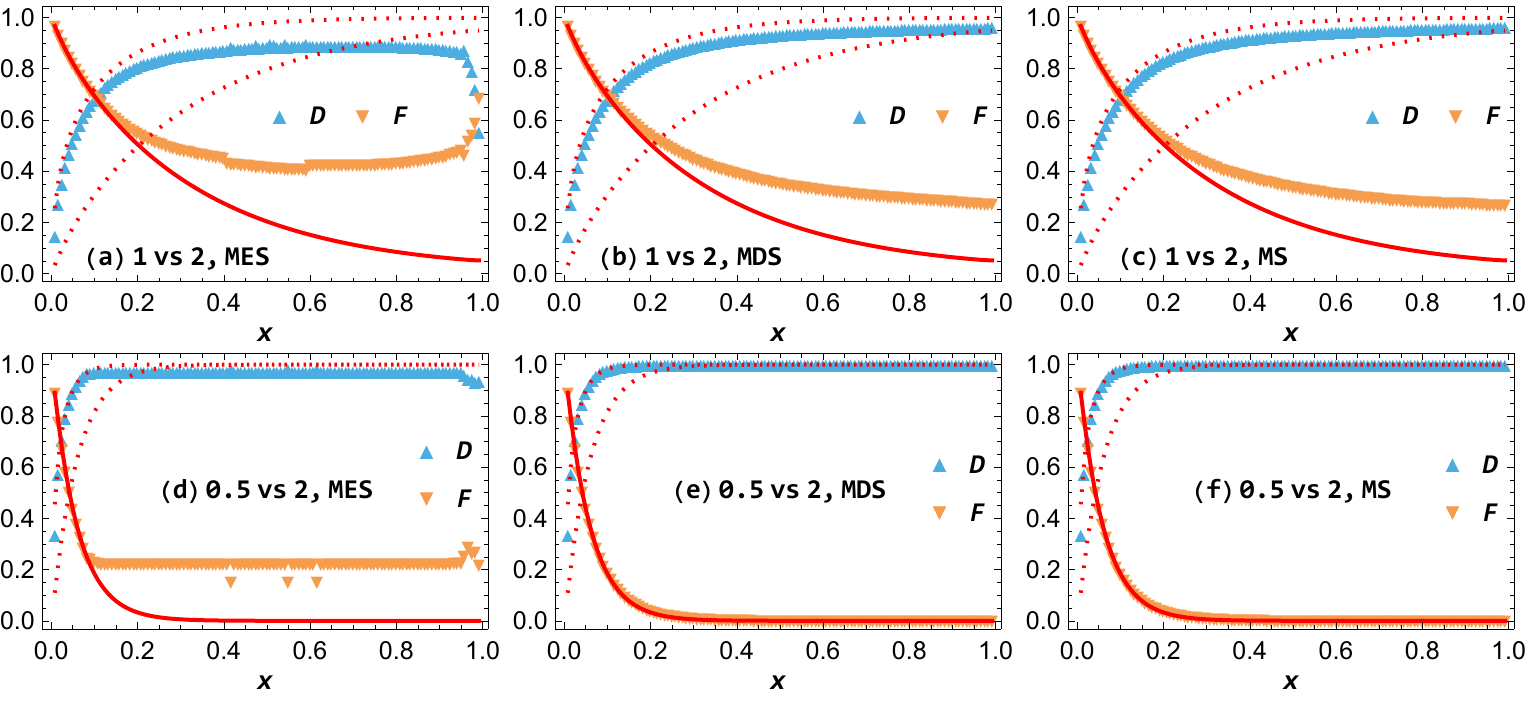}\\
  \caption{The presented examples showcase the subsystem trace distance $(D\equiv D(\r_{A,\l_1},\r_{A,\l_2}))$ and fidelity $(F\equiv F(\r_{A,\l_1},\r_{A,\l_2}))$ between two ground states of the Ising chain with varying transverse fields.
  These quantities are obtained using different strategies of the truncated correlation matrix method (depicted by symbols), while the exact fidelity results are represented by the red solid lines using formula (\ref{FrG1rG2}).
  Additionally, the red dotted lines indicate the lower and upper bounds for the trace distance derived from formula (\ref{uandlbounds}).
  In the top panels, we consider $\l_1=1$ and $\l_2=2$, while in the bottom panels, we have $\l_1=0.5$ and $\l_2=2$. The values of $L=120$ and $t=10$ are also set for this analysis.}
  \label{FigureGSIsingTDF}
\end{figure*}

Our aim is to calculate the approximate trace distance $D(\rho_{A,\lambda_1},\rho_{A,\lambda_2})$ and fidelity $F(\rho_{A,\lambda_1},\rho_{A,\lambda_2})$ for various values of $\lambda_1$ and $\lambda_2$. Examples are illustrated in Figure~\ref{FigureGSIsingTDF}. It is worth noting that all the ground state RDMs considered in this section correspond to small-entropy states, characterized by either finite or logarithmic-law von Neumann entropies. However, for two different small-entropy states, the noncommutativity trace norm (\ref{NCTN}) exhibits linear growth with respect to the subsystem size $\ell$. According to the conditions stated in (\ref{TheConditions}), this suggests that the truncation method is generally not effective in such cases (as shown in the top panels of Figure~\ref{FigureGSIsingTDF}). However, for cases where the two density matrices are nearly orthogonal, indicated by the presence of maximal canonical values of the correlation matrix difference approaching 2, the truncation method works well, as demonstrated in the bottom panels of Figure~\ref{FigureGSIsingTDF}.

\section{Examples of low-lying eigenstate RDMs in critical Ising chain} \label{sectioncriticalIsing}

As an intriguing application of the truncation method, we explore the subsystem trace distance and fidelity among several low-lying eigenstates in the critical Ising chain, i.e.\ the Hamiltonian (\ref{HIsing}) with $\l=1$. The critical Ising chain is known to be described by a conformal field theory, specifically the two-dimensional free fermion theory.

In our analysis, we focus on states corresponding to the low-lying eigenstates $|G\rangle$, $|\sigma\rangle$, $|\mu\rangle$, $|\psi\rangle$, $|\bar{\psi}\rangle$, and $|\varepsilon\rangle$ in the two-dimensional free fermion theory. The ground state is denoted by $|G\rangle$.
The states $|\sigma\rangle$, $|\mu\rangle$, $|\psi\rangle$, $|\bar{\psi}\rangle$, and $|\varepsilon\rangle$ correspond to primary operators $\sigma$, $\mu$, $\psi$, $\bar{\psi}$, and $\varepsilon$ within the framework of two-dimensional conformal field theory. The state-operator correspondence in two-dimensional conformal field theory establishes that these primary operators give rise to corresponding primary excited states. For a more comprehensive understanding of the correspondence between spin chain eigenstates and field theory states, further details can be found in references such as \cite{Berganza:2011mh}.

\begin{figure*}[tp]
  \centering
  \includegraphics[width=0.92\textwidth]{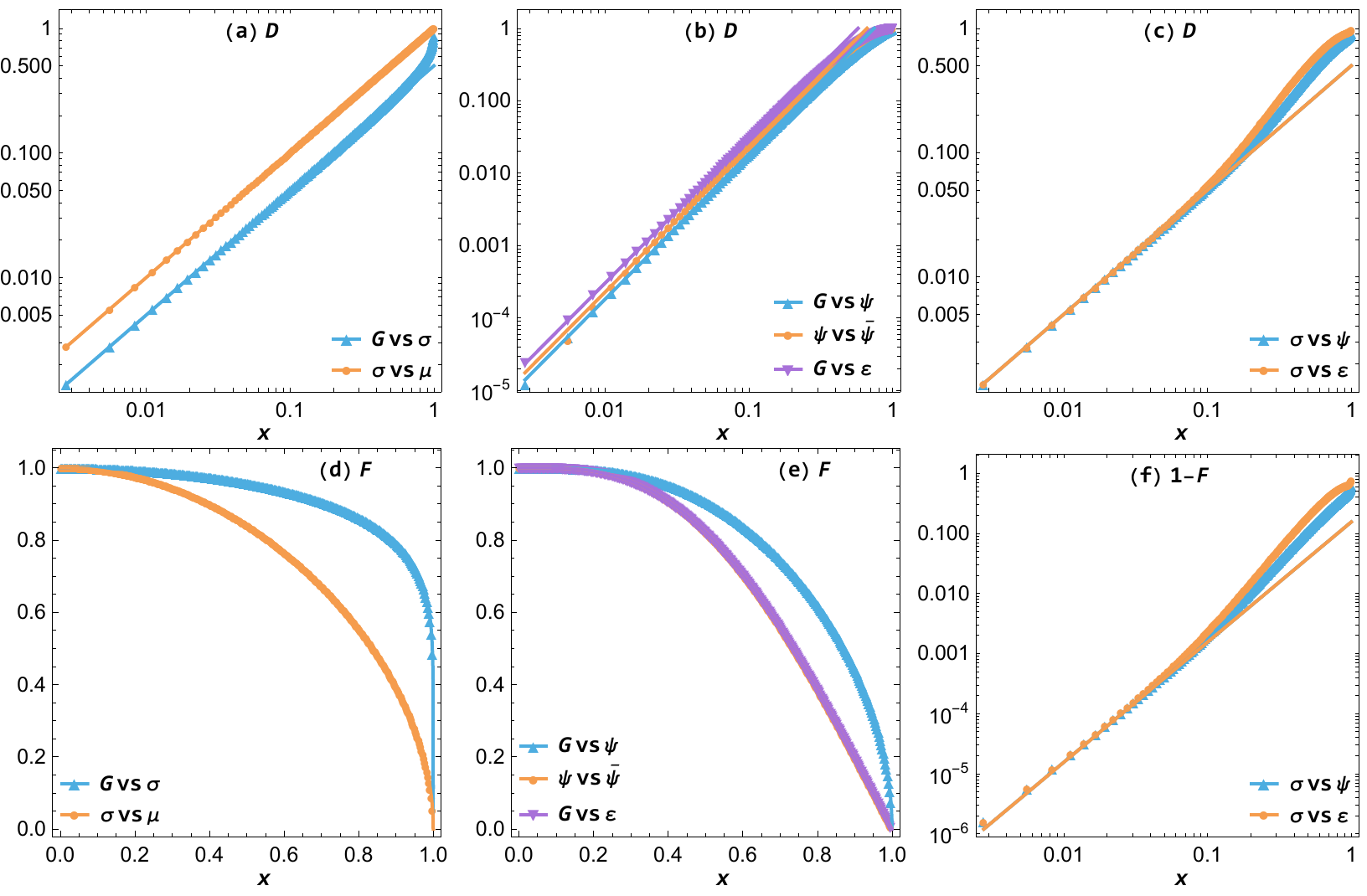}\\
  \caption{The solid lines represent the trace distance $(D)$ and fidelity $(F)$ between low-lying eigenstates in the two-dimensional free fermion theory. The symbols depict the corresponding approximate trace distance and fidelity obtained using the mixed strategy of the truncation strategy in the critical Ising chain.
  Note that in panel (e), the fidelity $F(\r_{A,\psi},\r_{A,\bar \psi})$ coincides with $F(\r_{A,G},\r_{A,\ve})$.
  For the Ising chain, we have chosen a system size of $L=360$ and a truncation number of $t=10$.}
  \label{FigureIsingcriticalTDF}
\end{figure*}

In \cite{Zhang:2019wqo,Zhang:2019itb}, there are results of trace distances
\bea
&& D(\r_{A,G},\r_{A,\s}) = D(\r_{A,G},\r_{A,\mu}) = \f{x}{2} + o(x), \nn\\
&& D(\r_{A,\s},\r_{A,\m}) = x, \nn\\
&& D(\r_{A,\s},\r_{A,\psi}) = D(\r_{A,\s},\r_{A,\bar\psi}) = D(\r_{A,\m},\r_{A,\psi}) \nn\\
&& \phantom{D(\r_{A,\s},\r_{A,\psi})}= D(\r_{A,\m},\r_{A,\bar\psi}) = \f{x}{2} + o(x), \nn\\
&& D(\r_{A,\s},\r_{A,\ve}) = D(\r_{A,\m},\r_{A,\ve}) = \f{x}{2} + o(x), \nn\\
&& D(\r_{A,G},\r_{A,\psi}) = D(\r_{A,G},\r_{A,\bar \psi}) = D(\r_{A,\psi},\r_{A,\ve}) \nn\\
&& \phantom{D(\r_{A,G},\r_{A,\psi})} = D(\r_{A,\bar \psi},\r_{A,\ve}) \approx 1.81 
x^2 + o(x^2), \nn\\
&& D(\r_{A,\psi},\r_{A,\bar \psi}) \approx 2.27 
x^2 + o(x^2), \nn\\
&& D(\r_{A,G},\r_{A,\ve}) \approx 3.02 
x^2 + o(x^2),
\eea
and fidelities
\bea
&& F(\r_{A,G},\r_{A,\s}) = F(\r_{A,G},\r_{A,\mu}) = \Big( \cos\f{\pi x}{2} \Big)^{\f18}, \nn\\
&& F(\r_{A,\s},\r_{A,\m}) = \Big( \cos\f{\pi x}{2} \Big)^{\f12}, \nn\\
&& F(\r_{A,\s},\r_{A,\psi}) = F(\r_{A,\s},\r_{A,\bar\psi}) = F(\r_{A,\m},\r_{A,\psi}) \nn\\
&& \phantom{F(\r_{A,\s},\r_{A,\psi})} = F(\r_{A,\m},\r_{A,\bar\psi}) = 1 - \f{\pi^2x^2}{64} + o(x^2), \nn\\
&& F(\r_{A,\s},\r_{A,\ve}) = F(\r_{A,\m},\r_{A,\ve}) = 1 - \f{\pi^2x^2}{64} + o(x^2), \nn\\
&& F(\r_{A,G},\r_{A,\psi}) = F(\r_{A,G},\r_{A,\bar \psi}) = F(\r_{A,\psi},\r_{A,\ve}) \\
&& \phantom{F(\r_{A,G},\r_{A,\psi})} = F(\r_{A,\bar \psi},\r_{A,\ve}) =
      \f{\G(\f{3+\csc\f{\pi x}{2}}{4})}
      {\G(\f{1+\csc\f{\pi x}{2}}{4})}
      \sqrt{2\sin(\pi x)}, \nn\\
&& F(\r_{A,\psi},\r_{A,\bar \psi}) = F(\r_{A,G},\r_{A,\ve}) =
      \f{\G^2(\f{3+\csc\f{\pi x}{2}}{4})}
      {\G^2(\f{1+\csc\f{\pi x}{2}}{4})}
      2\sin(\pi x). \nn
\eea
Note that the fidelity between the ground state and a general primary excited state in two-dimensional conformal field theory was initially derived in \cite{Lashkari:2014yva}. The calculations of the trace distance and fidelity in the critical Ising chain presented in \cite{Zhang:2019wqo,Zhang:2019itb} are exact, but they are limited to cases with very small subsystem sizes.
In our study, we have recalculated the trace distance and fidelity for significantly larger subsystem sizes by employing a mixed strategy of the truncation method. The results are depicted in figure~\ref{FigureIsingcriticalTDF}. Remarkably, there is excellent agreement between the approximate results obtained from the spin chain and the analytical results from field theory.

The success of the truncation method can be attributed to the fact that all relevant spin chain reduced density matrices are nearly commuting small-entropy states. This means that the non-commutativity trace norm of the two correlation matrices, as defined in (\ref{NCTN}), either approaches a constant or exhibits very slow growth with respect to the system size. Furthermore, the relevant RDMs are small entropy states, i.e.\ the von Neumann entropies of the RDMs are finite or follow a logarithmic law.

\section{Examples of low-lying eigenstate RDMs in half-filled XX chain} \label{sectionlowlyingXX}

We investigate the subsystem trace distance between eigenstates in the XX chain described by the Hamiltonian
\be
H = - \sum_{j=1}^L \Big[ \f14 ( \s_j^x\s_{j+1}^x + \s_j^y\s_{j+1}^y ) +\f\l2 \s_j^z \Big].
\ee
By applying the Jordan-Wigner transformation, the XX chain can be mapped to the free fermion theory with the Hamiltonian
\be
H = - \f12 \sum_{j=1}^L \Big[ a_j^\dag a_{j+1} + a_{j+1}^\dag a_j + \l(1-2 a_j^\dag a_j) \Big],
\ee
where the number of excited Dirac modes is a conserved quantity given by $N = \sum_{j=1}^L a^\dag_j a_j$.

To compute the subsystem trace distance, we have two methods at our disposal: the truncated canonicalized correlation matrix method, as discussed in Section~\ref{sectionTCCMM}, or the truncated diagonalized correlation matrix method, as introduced in Appendix~\ref{sectionTDCMM}.
It is worth noting that the classification of small-entropy states, which refers to states with finite or logarithmic-law von Neumann entropies, and large-entropy states, which correspond to states with volume-law von Neumann entropies, follows the same criteria as in the Ising chain.
The results and conclusions obtained for the XX chain are consistent with those of the Ising chain. The truncation method is effective for two small-entropy states that exhibit nearly commutative properties. However, when dealing with large-entropy states, the truncation method generally proves to be unsuccessful, except in cases where two special nearly orthogonal Gaussian RDMs are involved. It is important to note that these nearly orthogonal Gaussian states are characterized by the presence of eigenvalues of the correlation matrix difference that are close to $-1$ or $1$.

In this paper, we present the results of the trace distance and fidelity analysis for several low-lying eigenstates in the XX chain without a transverse field ($\lambda=0$), while omitting the detailed calculations. The ground state of this system exhibits a half-filled momentum space.
The critical XX chain, in the continuum limit, corresponds to the two-dimensional free compact boson theory, which is also recognized as a conformal field theory.
Within the XX chain, we focus on specific states, namely the ground state denoted by $|G\rangle$, current states represented by $|J\rangle$ and $|\bar{J}\rangle$, the state $|J\bar{J}\rangle$, and the vertex operator state denoted by $|\alpha, \bar{\alpha}\rangle$ defined as $|\alpha, \bar{\alpha}\rangle = V_{\alpha,\bar{\alpha}}|G\rangle$. It is worth noting that $|0,0\rangle$ is equivalent to $|G\rangle$.

\begin{figure*}[tp]
  \centering
  \includegraphics[width=0.92\textwidth]{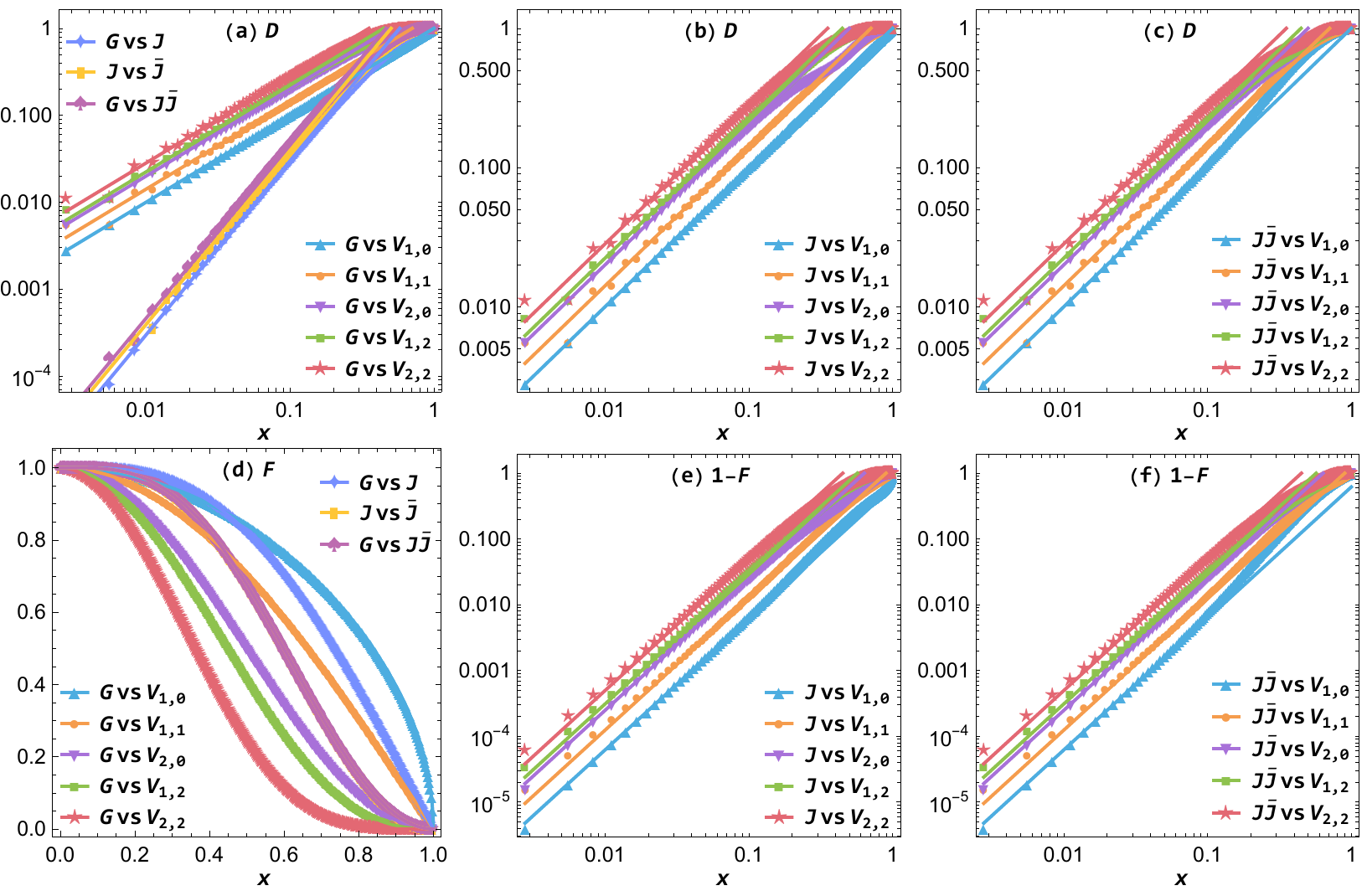}\\
  \caption{The solid lines represent the trace distance $(D)$ and fidelity $(F)$ between low-lying eigenstates in the two-dimensional free compact bosonic theory. The symbols depict the corresponding approximate trace distance and fidelity obtained using the mixed strategy of the truncation method in the half-filled XX chain.
  Note that in panel (d), the fidelity $F(\r_{A,J},\r_{A,\bar J})$ coincides with the fidelity $F(\r_{A,G},\r_{A,J\bar J})$.
  For the XX chain, we have chosen a system size of $L=360$ and a truncation number of $t=10$.}
  \label{FigureXXcriticalTDF}
\end{figure*}

In \cite{Zhang:2019wqo,Zhang:2019itb}, one can find the field theory results for the trace distance
\bea
&& D(\r_{A,\a,\bar\a},\r_{A,\a',\bar\a'}) = x, {{\rm~if~} (\a-\a')^2+(\bar\a-\bar\a')^2=1}, \nn\\
&& D(\r_{A,\a,\bar\a},\r_{A,\a',\bar\a'}) = \sqrt{(\a-\a')^2+(\bar\a-\bar\a')^2} x + o(x), \nn\\
&& D(\r_{A,J},\r_{A,\a,\bar\a}) = D(\r_{A,\bar J},\r_{A,\bar\a,\a}) = \sqrt{\a^2+\bar\a^2} x + o(x), \nn\\
&& D(\r_{A,J\bar J},\r_{A,\a,\bar\a}) = \sqrt{\a^2+\bar\a^2} x + o(x), \nn\\
&& D(\r_{A,G},\r_{A,J}) = D(\r_{A,G},\r_{A,\bar J}) = D(\r_{A,J},\r_{A,J\bar J}) \nn\\
&& \phantom{D(\r_{A,G},\r_{A,J})} = D(\r_{A,\bar J},\r_{A,J\bar J}) \approx 2.99 x^2 + o(x^2), \nn\\
&& D(\r_{A,J},\r_{A,\bar J}) \approx 3.94 x^2 + o(x^2), \nn\\
&& D(\r_{A,G},\r_{A,J\bar J}) \approx 4.63 x^2 + o(x^2),
\eea
and fidelity
\bea
&& F(\r_{A,\a,\bar\a},\r_{A,\a',\bar\a'}) = \Big( \cos\f{\pi x}{2} \Big)^{\f{(\a-\a')^2+(\bar\a-\bar\a')^2}{2}}, \nn\\
&& F(\r_{A,J},\r_{A,\a,\bar\a}) = F(\r_{A,\bar J},\r_{A,\bar\a,\a}) \nn\\
&& \phantom{F(\r_{A,J},\r_{A,\a,\bar\a})} = 1 - (\a^2+\bar\a^2) \f{\pi^2 x^2}{16} + o(x^2), \nn\\
&& F(\r_{A,J\bar J},\r_{A,\a,\bar\a}) = 1 - (\a^2+\bar\a^2) \f{\pi^2 x^2}{16} + o(x^2), \nn\\
&& F(\r_{A,G},\r_{A,J}) = F(\r_{A,G},\r_{A,\bar J}) = F(\r_{A,J},\r_{A,J\bar J}) \\
&& \phantom{F(\r_{A,G},\r_{A,J})} = F(\r_{A,\bar J},\r_{A,J\bar J}) =
      \f{\G^2(\f{3+\csc\f{\pi x}{2}}{4})}
      {\G^2(\f{1+\csc\f{\pi x}{2}}{4})}
      2\sin(\pi x), \nn\\
&& F(\r_{A,J},\r_{A,\bar J}) = F(\r_{A,G},\r_{A,J\bar J}) =
      \f{\G^4(\f{3+\csc\f{\pi x}{2}}{4})}
      {\G^4(\f{1+\csc\f{\pi x}{2}}{4})}
      4\sin^2(\pi x).\nn
\eea
The fidelity between the ground state and any primary excited state in two-dimensional conformal field theory was determined in \cite{Lashkari:2014yva}. In Figure~\ref{FigureXXcriticalTDF}, we present the results for the critical XX chain and the free compact bosonic theory. Once again, the truncation method exhibits remarkable effectiveness due to the near-commutativity of all pertinent reduced density matrices, which are all small-entropy states.

\section{Conclusion and discussions} \label{sectionConclusion}

In this paper, we have developed a truncation method for evaluating the trace distance between two fermionic Gaussian states. To verify the efficacy of the method, we examined various examples of RDMs in the Ising and XX chains. Our findings indicate that the method performs well under two distinct sets of conditions, as outlined in (\ref{TheConditions}).
The first condition pertains to the von Neumann entropies of the two states, which should be small.
By ``small,'' we mean that the von Neumann entropies of both states either are finite or follow a logarithmic law.
Additionally, the correlation matrices of the two states should exhibit near-commutation behavior.
The second condition is related to the near-orthogonality of the two states, as characterized by the correlation matrix difference having a maximal canonical value close to 2.
As an application of the truncation method, we have computed the subsystem trace distances among the ground state and the low-lying excited states in the critical Ising chain and the half-filled XX chain. The results obtained in Sections \ref{sectioncriticalIsing} and \ref{sectionlowlyingXX} demonstrate excellent agreement with the predictions derived from the corresponding two-dimensional conformal field theories. This represents a significant improvement over the calculations of subsystem trace distances presented in \cite{Zhang:2019wqo,Zhang:2019itb}, from around 10 sites to several hundreds sites.

The method we have introduced does not always yield satisfactory results, as demonstrated by the examples discussed in Sections \ref{sectionIsing} and \ref{sectionIsingGS}. By truncating the dimension of the Hilbert space and carefully selecting a limited number of effective modes, the truncation method can potentially be successful when applied to two states that share similar effective modes. Moreover, it is important to note that only a finite number of effective modes significantly contribute to the trace distance.
The condition of near-orthogonality outlined in (\ref{TheConditions}) pertains to specific cases that are relatively straightforward to comprehend. When the maximal canonical value of the correlation matrix difference approaches 2, the maximal difference strategy presented in Subsection \ref{MDSofTCCMM} can identify an effective mode associated with this maximal canonical value and provide an approximate trace distance value close to 1.
For more general scenarios, we anticipate that the truncation method will be effective when the two states possess nearly identical effective modes and only a limited number of these modes have major contributions to the trace distance.
These conditions correspond to the nearly commuting and small entropy conditions outlined in (\ref{TheConditions}), respectively.
In Section \ref{sectionIsing}, each unsuccessful example features at least one state with a von Neumann entropy that follows a volume law, thereby violating the small entropy condition. Similarly, in Section \ref{sectionIsingGS}, each unsuccessful example involves a noncommutativity trace norm of the two correlation matrices that scales linearly with the system size, violating the nearly commuting condition.
In this paper, we have summarized the conditions under which the truncation method can succeed based on the examples provided. However, it remains an intriguing area of research to gain a deeper understanding of how and why the method works, as well as the circumstances under which it may fail.

It is important to note that a finite truncation method is not always feasible for two Gaussian states with volume law entropies, as illustrated by an example in Appendix \ref{appendixToy}. However, there are instances of fermionic Gaussian states for which a finite truncation method is possible, yet the approach presented in this paper may fail. An example of such a case can be found in Section \ref{sectionIsingGS}, where we examined the trace distance between ground state reduced density matrices in Ising chains with different transverse fields.
An intriguing avenue for future research would be to generalize the truncation method introduced in this paper and apply it to a broader range of fermionic Gaussian states, as well as non-Gaussian states. It is worth noting that calculating fidelity and relative entropy is often easier than determining the trace distance. One possible approach to calculating the trace distance is as follows: identify a finite-dimensional subspace within the Hilbert space that minimize the fidelity or maximize the relative entropy and then compute the trace distance within that subspace. A similar truncation method could also be employed to compute other quantities in fermionic systems and spin chains, such as entanglement negativity \cite{Vidal:2002zz,Plenio:2005cwa} and reflected entropy \cite{Dutta:2019gen}.
We look forward to revisiting this problem in future investigations.

\section*{Acknowledgements}

We thank Markus Heyl and Reyhaneh Khaseh for reading a previous version of the draft and helpful comments.
We thank the anonymous referees for valuable comments and suggestions.
JZ acknowledges support from the National Natural Science Foundation of China (NSFC) through grant number 12205217.
MAR thanks CNPq and FAPERJ (grant number 210.354/2018) for partial support.

\appendix

\section{An example of orthogonal states with no possible truncation} \label{appendixToy}

For certain orthogonal or nearly orthogonal Gaussian states, it is not necessary for the canonical values of the correlation matrix difference to be nearly 2 in the scaling limit. In this appendix, we provide an example of two orthogonal Gaussian states in the scaling limit, where all the canonical values of the correlation matrix difference can be smaller than 2. In such cases, a finite truncation method for the trace distance does not exist.

\begin{figure}[tp]
  \centering
  \includegraphics[width=0.3\textwidth]{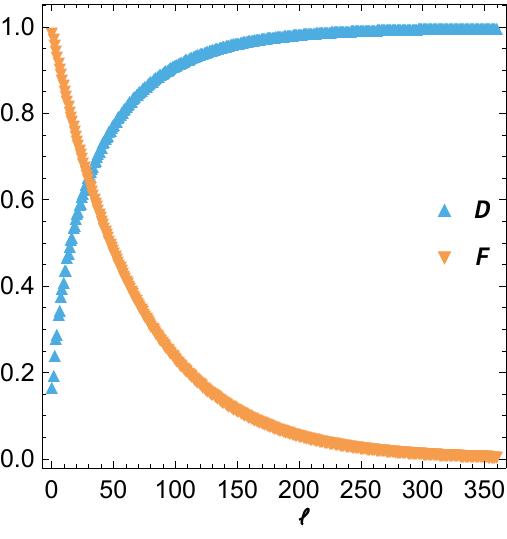}\\
  \caption{Trace distance and fidelity (\ref{ToyTDF}) computed between two toy model states (\ref{toyDM}) characterized by probabilities $p_1=\f13$ and $p_2=\f12$.}
  \label{FigureToy}
\end{figure}

We consider the Gaussian state density matrix in its simplest form, given by
\be \label{toyDM}
\r_p = \bigotimes_{j=1}^\ell \Big( \ba{cc} p & \\ & 1-p \ea \Big),
\ee
where $p\in(0,1)$.
The corresponding correlation matrix is
\be
\G_p =  \bigoplus_{j=1}^\ell \Big( \ba{cc}  & \ii (1-2p) \\ - \ii (1-2p) &  \ea \Big).
\ee
The von Neumann entropy of the density matrix follows a volume law
\be
S(\r_p) = \ell [ -p\log p - (1-p)\log(1-p) ].
\ee
The trace distance and fidelity can be easily obtained as
\bea \label{ToyTDF}
&& D(\r_{p_1},\r_{p_2}) = \f12 \sum_{i=0}^\ell C_\ell^i \big| p_1^i (1-p_1)^{\ell-i} - p_2^i (1-p_2)^{\ell-i} \big|, \nn\\
&& F(\r_{p_1},\r_{p_2}) = \big[ \sqrt{p_1p_2} + \sqrt{(1-p_1)(1-p_2)} \big]^\ell.
\eea
We provide an example with $p_1=\f13$ and $p_2=\f12$ in figure~\ref{FigureToy}.
Even when $p_1$ and $p_2$ are very close to each other, as long as $p_1\neq p_2$, we observe the following limits as $\ell\to+\inf$
\bea
&& \lim_{\ell\to+\inf} D(\r_{p_1},\r_{p_2}) = 1,\nn\\
&& \lim_{\ell\to+\inf} F(\r_{p_1},\r_{p_2}) = 0.
\eea
This indicates that the two states $\r_{p_1}$ and $\r_{p_2}$ with $p_1\neq p_2$ become orthogonal in the scaling limit $\ell\to+\inf$.
With finite truncation, it is impossible to approach this limit.

The correlation matrix difference is given by
\be
\G_{p_1}-\G_{p_2} = \bigoplus_{j=1}^\ell \Big( \ba{cc}  & -2 \ii (p_1-p_2) \\ 2 \ii (p_1-p_2) &  \ea \Big).
\ee
All the canonical values of the correlation matrix difference are equal to $2|p_1-p_2|$. However, this value is only equal to 2 in the special trivial cases when $p_1=1$ and $p_2=0$, or when $p_1=0$ and $p_2=1$.
This observation confirms that in the scaling limit, it is not necessary for the canonical values of the correlation matrix difference to be nearly 2 for certain orthogonal or nearly orthogonal Gaussian states.

\section{Truncated diagonalized correlation matrix method} \label{sectionTDCMM}

In the context of Gaussian states in a free fermionic theory, where the number of excited Dirac modes is conserved, a specialized version of the truncation method called the truncated diagonalized correlation matrix method can be employed.

In this appendix, we begin by reviewing the diagonalized correlation matrix method described in \cite{Zhang:2022tgu}, and subsequently extend it to the truncated diagonalized correlation matrix method. The overall procedure is similar to that outlined in section~\ref{sectionTCCMM}, with the distinction being the transformation from canonicalization of purely imaginary anti-symmetric correlation matrices to the diagonalization of Hermitian correlation matrices. The truncated diagonalized correlation matrix method proves to be advantageous, as the diagonalization of an $\ell\times\ell$ Hermitian matrix is generally easier compared to the canonicalization of the $2\ell\times2\ell$ purely imaginary anti-symmetric matrix.

\subsection{Diagonalized correlation matrix method}

We consider a system of $\ell$ spinless fermions denoted by $a_j$ and $a_j^\dag$, where $j = 1, 2, \ldots, \ell$. In the context of Gaussian states, these states can be characterized by the correlation matrix $C$, defined as
\be \label{Cdef}
C_{j_1 j_2} = \lag a_{j_1}^\dag a_{j_2} \rag, ~ j_1, j_2 = 1, 2, \ldots, \ell.
\ee

The trace distance and fidelity between two Gaussian states can be computed using the diagonalized correlation matrix method described in \cite{Zhang:2022tgu}. The correlation matrix $C$ satisfies the eigenvalue equation
\be
C u_j = \m_j u_j, ~ j=1,2,\ldots,\ell,
\ee
where $\m_j$ are real eigenvalues in the range $[0,1]$, and $u_j$ are the corresponding eigenvectors. We define the unitary matrix as
\be
U \equiv (u_1,u_2,\ldots,u_\ell),
\ee
where each eigenvector is represented as a column vector. The correlation matrix $C$ can be diagonalized as
\be \label{UdagCAKU}
U^\dag C U = \diag(\m_1,\m_2,\ldots,\m_\ell).
\ee
The density matrix can be expressed in terms of the modular Hamiltonian, as shown in \cite{Cheong:2002ukf,Peschel:2002jhw}, given by
\be
\r = \det(1-C) \exp\Big(-\sum_{j_1,j_2=1}^\ell M_{j_1j_2}a_{j_1}^\dag a_{j_2}\Big),
\ee
where the matrix $M$ is defined as
\be
M = \log(C^{-1} - 1).
\ee
Notably, the matrix $M$ is also diagonal under the same basis, satisfying
\bea
&& U^\dag M U = \diag(\n_1,\n_2,\ldots,\n_\ell), \nn\\
&& \n_j = \log(\m_j^{-1}-1), ~ j=1,2,\ldots,\ell.
\eea

We introduce the effective Dirac modes as
\be
\td a_j = \sum_{j'=1}^\ell U^\dag_{jj'} a_{j'}, ~
\td a_j^\dag = \sum_{j'=1}^\ell U_{j'j} a_{j'}^\dag.
\ee
Using these modes, the density matrix can be expressed as
\bea \label{rhoofdcmm}
&& \r  = \Big[ \prod_{j=1}^\ell(1-\m_j) \Big] \exp
      \Big( - \sum_{j=1}^\ell \n_j \td a_j^\dag \td a_j \Big) \nn\\
&& \phantom{\r}
    = \prod_{j=1}^\ell [ (1-\m_j) + (2\m_j-1) \td a_j^\dag \td a_j ].
\eea
To calculate the trace distance and fidelity, we employ the explicit form of the density matrices. For the fidelity calculation, it is convenient to use the square root of the density matrix:
\be
\sqrt{\r } = \prod_{j=1}^\ell [ \sqrt{1-\m_j} + ( \sqrt{\m_j} - \sqrt{1-\m_j} ) \td a_j^\dag \td a_j ].
\ee
To verify this formula, one can compare $\sqrt{\r}^2$ with $\r$ in equation (\ref{rhoofdcmm}).

\subsection{Truncated diagonalized correlation matrix method}

We perform a truncation of the Hilbert space by selecting a limited number of effective Dirac modes based on three different strategies.

\subsubsection{Maximal entropy strategy}

For the correlation matrix $C$ given by Eq. (\ref{UdagCAKU}), the corresponding density matrix $\rho$ can be transformed through a similarity transformation to the following form
\be
\r  \cong \bigotimes_{j=1}^\ell \lt( \ba{cc} 1-\m_j & \\ & \m_j \ea \rt).
\ee
The von Neumann entropy of $\rho$ is obtained by summing the Shannon entropy of each effective probability distribution $\{1-\m_j, \m_j\}$ with $j=1,2,\cdots,\ell$
\be
S(\r)  = \sum_{j=1}^\ell [ -(1-\m_j)\log(1-\m_j) -\m_j\log\m_j ].
\ee
For a low-rank state $\rho$, it is possible to truncate the effective probability distribution by selecting a few cases where $\m_j$ is closest to $\frac{1}{2}$. We follow this idea and truncate two density matrices to calculate their trace distance.

We consider two states $\rho_1$ and $\rho_2$ with correlation matrices $C_1$ and $C_2$. The matrix $C_1$ has $\ell$ pairs of eigenvalues and eigenvectors $(\m_j,u_j)$ with $j=1,2,\cdots,\ell$, and the matrix $C_2$ has $\ell$ pairs of eigenvalues and eigenvectors $(\m_j,u_j)$ with $j=\ell+1,\ell+2,\cdots,2\ell$. We sort the $2\ell$ pairs of eigenvalues and eigenvectors $(\m_j,u_j)$ with $j=1,2,\cdots,2\ell$ based on the values of $|\m_j-\frac{1}{2}|$, from the smallest to the largest. We choose the first $t$ vectors and rename them as $v_i$ with $i=1,2,\cdots,t$. These $t$ selected modes have the greatest contributions to the sum of the von Neumann entropies
\be
S(\r_1) + S(\r_2)  = \sum_{j=1}^{2\ell} [ -(1-\m_j)\log(1-\m_j) -\m_j\log\m_j ].
\ee

The vectors $v_i$ with $i=1,2,\cdots,t$ are generally not orthogonal and may not be independent. To orthogonalize these vectors, we introduce a $t \times t$ matrix $V$ with entries:
\be
V_{i_1i_2} \equiv v_{i_1}^\dag v_{i_2}, ~ i_1,i_2=1,2,\cdots,t.
\ee
We sort the $t$ pairs of eigenvalues and eigenvectors $(\a_i,a_i)$ with $i=1,2,\cdots,t$ of $V$ in descending order of $\a_i$, discarding those smaller than a cutoff, e.g., $10^{-9}$. After discarding, we obtain $\td \ell$ sets of eigenvalues and eigenvectors $(\a_i,a_i)$ with $i=1,2,\cdots,\td\ell$. From these, we define $\td\ell$ orthonormal vectors
\be
\td u_i \equiv \f{1}{\sqrt{\a_i}} \sum_{i'=1}^t [a_i]_{i'} v_{i'} , ~ i=1,2,\cdots,\td\ell.
\ee
Next, we define the $\ell\times\td\ell$ matrix:
\be
\td U \equiv ( \td u_1, \td u_2,\cdots, \td u_{\td\ell} ),
\ee
and the $\td\ell \times \td\ell$ truncated correlation matrices
\be
\td C_1 \equiv \td U^\dag C_1 \td U, ~
\td C_2 \equiv \td U^\dag C_2 \td U.
\ee

Using the diagonalized correlation matrix method mentioned in the previous subsection, we construct the $2^{\td\ell}\times 2^{\td\ell}$ truncated density matrices $\td \rho_1$ and $\td \rho_2$ from the $\td\ell\times\td\ell$ truncated correlation matrices $\td C_1$ and $\td C_2$. With these truncated density matrices $\td \rho_1$ and $\td \rho_2$, we calculate the approximate trace distance and fidelity:
\bea
&& \td D_{t,0}(\r_1,\r_2) \equiv D(\td \r_1,\td \r_2), \\
&& \td F_{t,0}(\r_1,\r_2) \equiv F(\td \r_1,\td \r_2).
\eea

\subsubsection{Maximal difference strategy}

To compare two states $\rho_1$ and $\rho_2$, we select $t < \ell$ eigenvectors $u_1, u_2, \cdots, u_t$ of the correlation matrix difference $C_1-C_2$ that correspond to the $t$ largest absolute eigenvalues. We then define the $\ell\times t$ matrix as
\be
\td U \equiv (u_1, u_2, \cdots ,u_t).
\ee
The truncated correlation matrices are defined as
\be
\td C_1 \equiv \td U^\dag C_1 \td U, ~
\td C_2 \equiv \td U^\dag C_2 \td U.
\ee

Using the diagonalized correlation matrix method, we construct the $2^t\times 2^t$ truncated density matrices $\td \rho_1$ and $\td \rho_2$ from the $t\times t$ truncated correlation matrices $\td C_1$ and $\td C_2$. With these truncated density matrices $\td \rho_1$ and $\td \rho_2$, we calculate the approximate trace distance and fidelity:
\bea
&& \td D_{0,t}(\r_1,\r_2) \equiv D(\td \r_1,\td \r_2), \\
&& \td F_{0,t}(\r_1,\r_2) \equiv F(\td \r_1,\td \r_2).
\eea

\subsubsection{Mixed strategy}

We also employ a strategy that combines elements of both the maximal entropy strategy and the maximal difference strategy. For a fixed value of $t = t_1 + t_2$, we first obtain $t_1$ vectors $v_i$ with indices $i = 1, 2, \ldots, t_1$ using the maximal entropy strategy, and then we obtain $t_2$ vectors $v_i$ with indices $i = t_1 + 1, t_1 + 2, \ldots, t$ using the maximal difference strategy.
Next, we perform orthonormalization on the vectors $v_i$ with indices $i = 1, 2, \ldots, t$, resulting in a set of orthonormal vectors $\td u_i$ with indices $i = 1, 2, \ldots, \td\ell$, where $\td\ell$ is constrained to be less than or equal to $t$.

We define the $\ell\times\td\ell$ matrix as
\be
\td U \equiv ( \td u_1, \td u_2,\cdots, \td u_{\td\ell} ),
\ee
Additionally, we define the $\td\ell \times \td\ell$ truncated correlation matrices as
\be
\td C_1 \equiv \td U^\dag C_1 \td U, ~
\td C_2 \equiv \td U^\dag C_2 \td U.
\ee
Using the diagonalized correlation matrix method, we construct the $2^{\td\ell}\times 2^{\td\ell}$ truncated density matrices $\td \r_1$ and $\td \r_2$ from the $\td\ell\times\td\ell$ truncated correlation matrices $\td C_1$ and $\td C_2$.

To evaluate the approximate trace distance and fidelity, we calculate
\bea
&& \td D_{t_1,t_2}(\r_1,\r_2) \equiv D(\td \r_1,\td \r_2), \\
&& \td F_{t_1,t_2}(\r_1,\r_2) \equiv F(\td \r_1,\td \r_2).
\eea
Finally, we obtain estimates of the trace distance and fidelity using the mixed strategy of the truncation method
\bea
&& \td D_{t}(\r_1,\r_2) \equiv \max_{0 \leq t_1 \leq t} \td D_{t_1,t-t_1}(\r_1,\r_2), \\
&& \td F_{t}(\r_1,\r_2) \equiv \min_{0 \leq t_1 \leq t} \td F_{t_1,t-t_1}(\r_1,\r_2).
\eea


\providecommand{\href}[2]{#2}\begingroup\raggedright\endgroup

\end{document}